\documentclass[10pt]{article}

\usepackage{natbib}
\usepackage{amsmath}
\usepackage{txfonts}
\usepackage{graphicx}
\usepackage[small]{caption}

\newcommand{\change}[1]{{#1}}

\newcommand{\ion}[2]{{\text{{\sc #1}\,{\sc #2}}}}

\newcommand{\pot}[2]{#1 \times 10^{#2}}

\newcommand{\apj}{{\rm ApJ }}
\newcommand{\apjs}{{\rm ApJS }}
\newcommand{\apjl}{{\rm ApJL }}
\newcommand{\aap}{{\rm A\&A }}
\newcommand{\apss}{{\rm Astrophysics and Space Science}}
\newcommand{\mnras}{{\rm MNRAS }}

\title{Signals From the Epoch of \\Cosmological Recombination \\ ({\it Karl Schwarzschild Lecture})}

\author{R.A. Sunyaev$^{1,2}$ and J. Chluba$^{3,1}$
\\[2mm]
{\scriptsize $^1$ Max-Planck-Institut f\"ur Astrophysik, Karl-Schwarzschild-Str. 1,
85741 Garching, Germany }
\\ {\scriptsize $^2$ Space Research Institute, Russian Academy of Sciences, Profsoyuznaya 84/32,
117997 Moscow, Russia }
\\
{\scriptsize $^3$ Canadian Institute for Theoretical Astrophysics, 60 St. George Street,
Toronto, ON M5S 3H8, Canada}
}

\begin{document}

\maketitle

\begin{abstract} 
  The physical ingredients to describe the epoch of cosmological recombination
  are amazingly simple and well-understood. This fact allows us to take into
  account a very large variety of physical processes, still finding potentially
  measurable consequences for the {\it energy spectrum} and {\it temperature anisotropies} of the Cosmic Microwave Background (CMB).
  In this contribution we \change{provide a short historical overview in connection with the cosmological recombination epoch and its connection to the CMB. Also we highlight some of the detailed physics that were studied over the past few years in the context of the cosmological recombination of {\it hydrogen} and {\it helium}}. 

  The impact of these considerations is two-fold:
  (i) the associated release of photons during this epoch leads to interesting
  and {\it unique deviations} of the Cosmic Microwave Background (CMB) energy
  spectrum {\it from a perfect blackbody}, which, in particular at decimeter
  wavelength and the Wien part of the CMB spectrum, may become observable in the near future.
 Despite the fact that the abundance of helium is rather small, it \change{still
 contributes a sizeable amount of photons to the full recombination spectrum,
 leading to additional distinct spectral features.}
    Observing the spectral distortions from the epochs of hydrogen
    and helium recombination, in principle would provide an additional way to
    determine some of the key parameters of the Universe (e.g. the specific
    entropy, the CMB monopole temperature and the pre-stellar abundance of
    helium).
  Also it permits us to confront our detailed understanding of the
  recombination process with {\it direct observational evidence}.
  In this contribution we illustrate how the theoretical {\it spectral
    template} of the cosmological recombination spectrum may be utilized for
  this purpose.
 \change{We also show that because hydrogen and helium recombine at very different epochs it is possible to address questions related to the {\it thermal history} of our Universe. In particular the cosmological recombination radiation may allow us to distinguish between Compton $y$-distortions that were created by energy release {\it before} or {\it after} the recombination of the Universe finished.}

  (ii) with the advent of high precision CMB data, e.g. as will be available
  using the {\sc Planck} Surveyor or {\sc CMBpol}, a very accurate theoretical
  understanding of the {\it ionization history} of the Universe becomes
  necessary for the interpretation of the CMB temperature and polarization
  anisotropies.
  Here we show that the uncertainty in the ionization history due to several
  processes, which until now were not taken in to account in the standard
  recombination code {\sc Recfast}, \change{reaches the percent level}.
In particular $\ion{He}{ii}\rightarrow\ion{He}{i}$-recombination occurs
 significantly faster because of the presence of a tiny fraction of neutral
 hydrogen at $z\lesssim 2400$.
\change{Also recently it was demonstrated that in the case of \ion{H}{i} Lyman $\alpha$ photons the {\it time-dependence} of the emission process and the {\it asymmetry} between the emission and absorption profile cannot be ignored.}
  However, it is indeed surprising how {\it inert} the cosmological
  recombination history is even at percent-level accuracy.
  Observing the cosmological recombination spectrum should in principle allow
    us to directly check this conclusion, which until now is purely
    theoretical. Also it may allow to {\it reconstruct the ionization history}
    using observational data.

\end{abstract}


\section{Introduction}
\label{RS:sec:Intro}
The Gunn-Peterson effect demonstrated clearly that intergalactic gas is strongly ionized in our vicinity till at least redshift $z \sim  6.5$. We are sure that at very high redshifts $z \gg 1000$ the CMB temperature was so high that hydrogen in the primordial matter should be completely ionized
(see Fig 1). 
Today we have no doubts that Universe was practically neutral at redshifts  $20 \lesssim z \lesssim 1000$. The periods of {\it reionization} (connected with formation of first stars and enormously strong release of UV-radiation) and of {\it cosmological hydrogen recombination} at redshift $\sim1000$ are of special importance for modern cosmology because they permit us to collect a lot of information about {\it history}, {\it structure} and {\it key parameters} of our Universe.

\subsection*{What is so beautiful about cosmological recombination?}
\label{RS:sec:Intro1}
\begin{figure}
\centering 
\includegraphics[width=0.85\columnwidth]{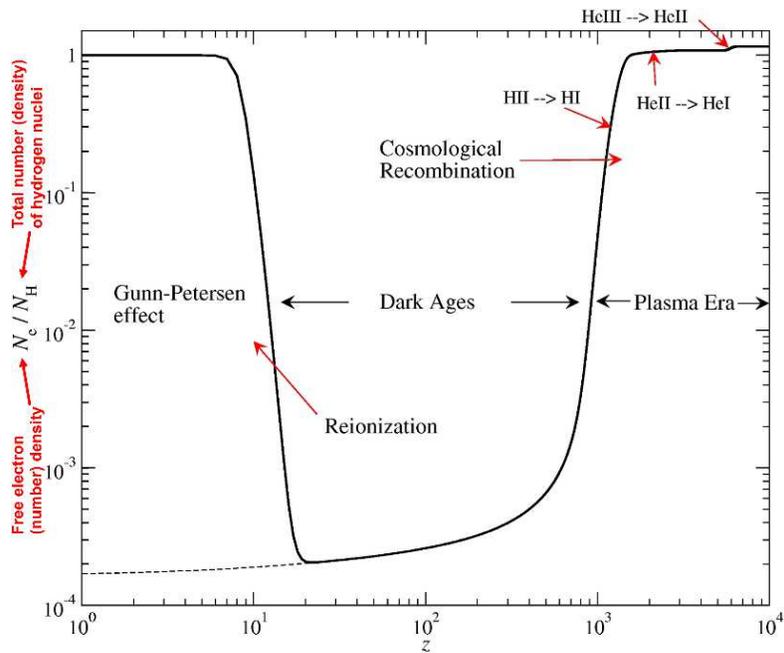}
\caption{Sketch of the cosmological ionization history as a function of redshift $z$. 
At high redshift the Universe was completely ionized. As it expanded and cooled down in went through several stages of recombination, starting with $\ion{He}{iii}\rightarrow \ion{He}{ii}$ recombination ($z\sim 7000$), $\ion{He}{ii}\rightarrow \ion{He}{i}$ recombination ($z\sim 2500$), and ending with the recombination of hydrogen ($z\sim 1000$). At low redshift ($z\lesssim 10$) the Universe eventually gets re-ionized by the first sources of radiation that appear in the Universe.
}
\label{fig:ion_hist}
\end{figure}

Within the cosmological concordance model the physical environment during the
epoch of cosmological recombination (redshifts $500 \lesssim z\lesssim 2000$
for hydrogen, $1600 \lesssim z\lesssim 3500$ for
\ion{He}{ii}$\rightarrow$\ion{He}{i} and $5000 \lesssim z\lesssim 8000$ for
\ion{He}{iii}$\rightarrow$\ion{He}{ii} recombination; also see Fig.~\ref{fig:ion_hist}) is extremely simple: the
Universe is homogeneous and isotropic, globally neutral and is expanding at a
rate that can be computed knowing a small set of cosmological parameters.
%
%
The baryonic matter component is dominated by hydrogen ($\sim 76\%$) and
helium ($\sim 24\%$), with negligibly small traces of other light
elements, such as deuterium and lithium, and it is continuously exposed to a
bath of isotropic blackbody radiation, which contains roughly $1.6\times 10^9$
photons per baryon.

\change{At redshift $z\sim 1400$ the electron number density in the Universe was close to $N_{\rm e} \sim 500\,{\rm cm}^{-3}$, a value that  is not very far from the densities of many compact \ion{H}{ii} regions in our Galaxy.
However, what makes the situation drastically different from the one in ionized nebulae is the {\it ambient bath of CMB photons} with the same temperature as electrons, $T_{\rm e}= T_{\gamma}\sim 3815\,$K, and the huge photon number density $N_\gamma\sim 1.1 \times 10^{12} {\rm cm}^{-3}$. 
In contrast to \ion{H}{ii} regions, radiative processes (instead of collisional processes and the interaction with strongly diluted stellar UV radiation spectrum) are most important. Furthermore, there are no heavy elements and dust. 
It is these conditions that make stimulated radiative processes, photoabsorption and ionization play an especially important role during hydrogen recombination. 
Another principle difference is the transition from problems with a spatial boundary in \ion{H}{ii} regions to the practically uniform Universe without boundaries.  Therefore the evolution of radiation in the expanding Universe is connected with a {\it time-dependent} rather than a {\it spatial} problem . 
}

These initially simple and very unique settings in principle allow us to
predict the {\it ionization history} of the Universe and the {\it cosmological
  recombination spectrum} (see Sects.~\ref{RS:sec:spectrum_all}) with extremely high accuracy, where the limitations
are mainly set by our understanding of the {\it atomic processes} and
associated transition rates.
In particular for neutral helium our knowledge is still rather poor.
  Only very recently highly accurate and user-friendly tables for the main
  transitions and energies of levels with $n\leq10$ have been published
  \citep{Drake2007}, but there is no principle difficulty in extending these
  to larger $n$ \citep{RS_BeigVain}. Also the data regarding the photo-ionization cross sections of
  neutral helium should be updated and extended.

In any case, it is this simplicity that offers us the possibility to enter a rich field of physical processes and to challenge our understanding of
atomic physics, radiative transfer and cosmology, eventually leading to a {\it
  beautiful} variety of potentially observable effects \change{in connection with the CMB radiation}.

\subsection*{What is so special about cosmological recombination?}
\label{RS:sec:Intro2}
%
The main reason for {the described} simplicity is the {\it extremely large specific
entropy} and the {\it slow expansion} of our Universe.
Because of the huge number of CMB photons, the free electrons are tightly coupled
to the radiation field due to tiny energy exchange during {\it Compton
    scattering} off thermal electrons until rather low redshifts, such that
during recombination the thermodynamic temperature of electrons is equal to
the CMB blackbody temperature with very high precision.
\change{Without this strong interaction between photons and electrons the temperature of the electrons would scale like $T_{\rm e}\propto (1+z)^2$, while the temperature of the photon field drops like $T_\gamma\propto (1+z)$.}
In addition, the very fast {\it Coulomb interaction} and atom-ion
  collisions allows to maintain full thermodynamic equilibrium among the
electrons, ions and neutral atoms down to $z\sim 150$ \citep{RS_Zeldovich68}.
\change{It is only below this redshift that the matter temperature starts to drop faster than the radiation temperature, a fact that is also very important in connection with the 21 cm signals coming from high redshift  before the Universe got reionized at $z\sim 10$ \citep{Madau1997}.}
Furthermore, processes in the baryonic sector cannot severely affect any of the
radiation properties, down to redshift where the first stars and galaxies
appear,
\change{so that as mentioned above} the atomic rates are largely dominated by radiative
processes, including {\it stimulated recombination}, {\it induced emission}
and absorption of photons.
On the other hand, the slow expansion of the Universe allows us to consider
the evolution of the atomic species along a sequence of {\it quasi-stationary}
stages, where the populations of the levels are nearly in full equilibrium
with the radiation field, but only subsequently and very slowly drop out of
equilibrium, finally leading to {\it recombination} and the {\it release of
  additional photons} in uncompensated bound-bound and free-bound transitions.
%

%
\begin{figure}
\centering 
\includegraphics[width=0.65\columnwidth]{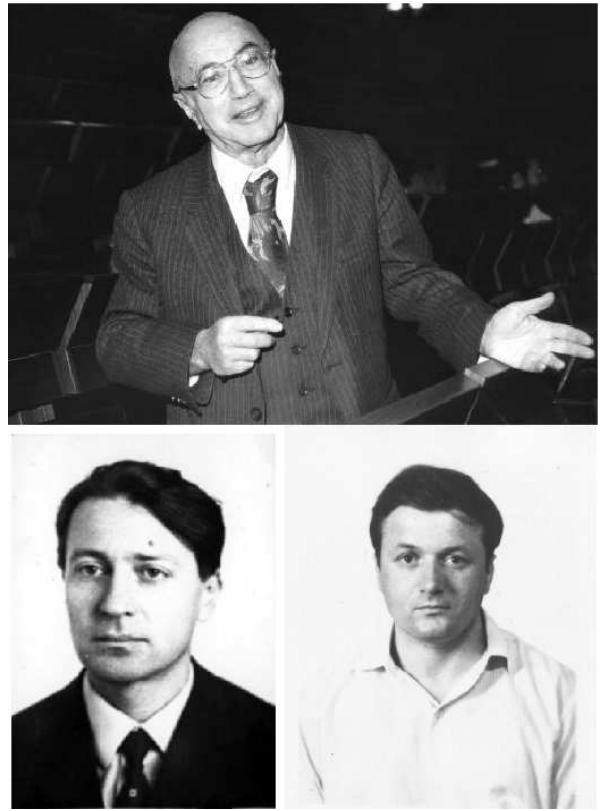}
\caption{Yakov B. Zeldovich (top), Vladimir Kurt (lower left) and RS (lower right).}
\label{RS:fig:ZKS}
\end{figure}

\begin{figure}
\centering 
\includegraphics[width=0.9\columnwidth]{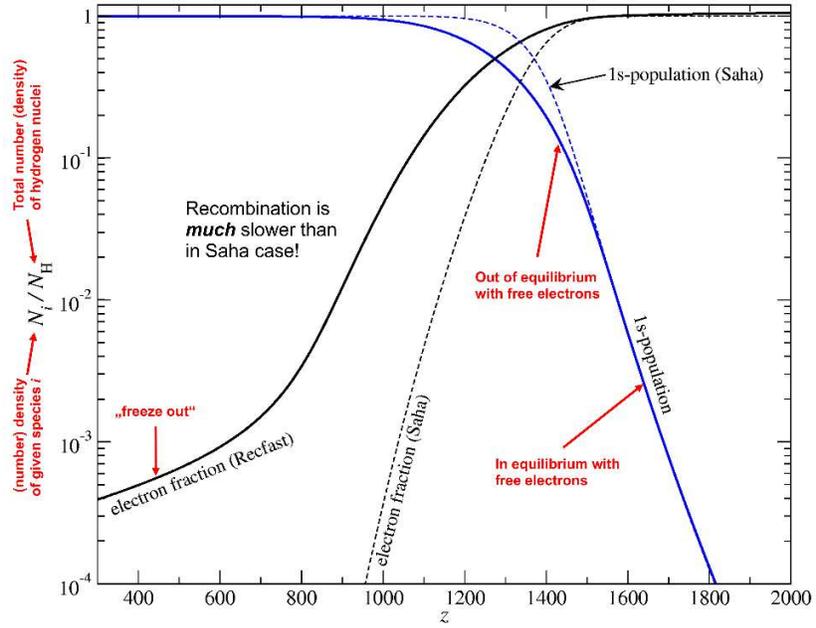}
\caption{Illustration of the difference in the hydrogen recombination history in comparison with the Saha case. The recombination of hydrogen in the Universe is strongly delayed due to the 'bottleneck' created in the Lyman $\alpha$ resonance and the slow 2s-1s two-photon transition. }
\label{fig:Xe.Saha}
\end{figure}

\begin{figure}
\centering 
\includegraphics[width=0.3\columnwidth]{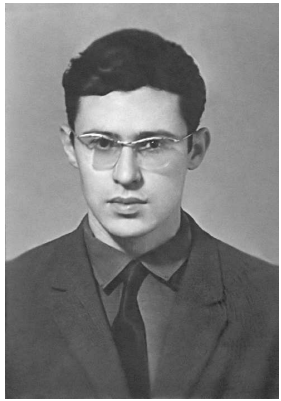}
\caption{Viktor Dubrovich.}
\label{RS:fig:VD}
\end{figure}
\subsection*{Brief historical overview {for hydrogen recombination}}
\label{RS:sec:history}
%
It was realized at the end of the 60's \citep{RS_Zeldovich68, RS_Peebles68},
that during the epoch of cosmological hydrogen recombination (typical
redshifts $800\lesssim z \lesssim 1600$) any direct recombination of electrons
to the ground state of hydrogen is immediately followed by the ionization of a
neighboring neutral atom due to re-absorption of the newly released
Lyman-continuum photon.
In addition, because of the enormous difference in the $2{\rm
  p}\leftrightarrow 1{\rm s}$ dipole transition rate and the Hubble expansion
{rate}, photons emitted close to the center of the Lyman-$\alpha$
line scatter $\sim 10^7-10^8$ times before they can finally escape further
interaction with the medium and thereby permit a successful settling
  of electrons in the 1s-level.
  It is due to these very peculiar circumstances that the $2{\rm
    s}\leftrightarrow 1{\rm s}$-two-photon decay process (transition rate $A_{\rm 2s1s}\sim 8.22\,{\rm s^{-1}}$), being about 8
  orders of magnitude slower than the Lyman-$\alpha$ resonance transition, is
  able to substantially control the dynamics of cosmological hydrogen
  recombination \citep{RS_Zeldovich68, RS_Peebles68},
  allowing about 57\% of all hydrogen atoms in the Universe to recombine at
  redshift $z\lesssim 1400$ through this channel \citep{RS_Chluba2006b}.

{Shortly afterwards \citep{RS_Sunyaev1970, RS_Peebles1970} it became
clear that the ionization history is one of the key ingredients for the
theoretical predictions of the Cosmic Microwave Background (CMB) temperature
and polarization anisotropies.} Today these tiny directional variations of the
CMB temperature ($\Delta T/T_0\sim 10^{-5}$) around the mean value
$T_0=2.725\pm 0.001\,$K \citep{RS_Fixsen2002} have been observed for the whole
sky using the {\sc Cobe}
%
%
and {\sc Wmap}
%
%
satellites, beyond doubt with great success.
The high quality data coming from balloon-borne and ground-based CMB
experiments ({\sc Boomerang, Maxima, Archeops, Cbi, Dasi} and {\sc Vsa} etc.)
today certainly provides one of the major pillars for the {\it cosmological
concordance model} \citep{Bahcall1999, RS_Bennett2003}. 
\change{Very recently the {\sc Planck} Surveyor was successfully launched and is now on its way to the L2 point, from which it will start observing the CMB with unprecedented precision very soon, further helping to establish the {\it era of precision cosmology}.}

\begin{figure}
\centering 
\includegraphics[width=0.98\columnwidth]{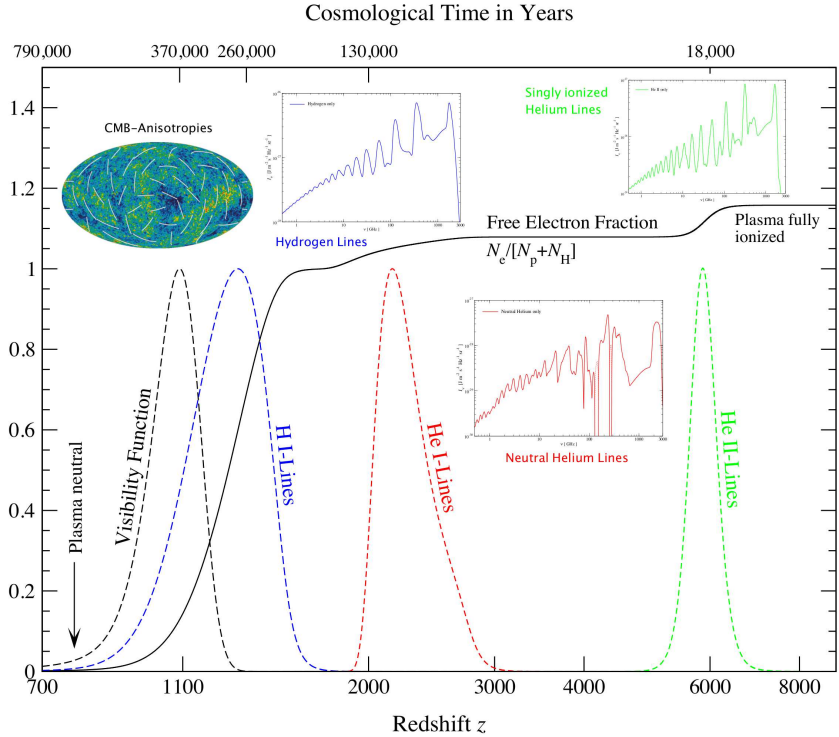}
\caption{Ionization history of the Universe (solid black curve) and the origin of different CMB signals \change{(dashed lines and inlays)}. 
  The observed temperature anisotropies in the CMB temperature are created close
  to the maximum of the Thomson visibility function around $z\sim 1089$,
  whereas the direct information carried by the photons in the cosmological
  hydrogen recombination spectrum is from slightly earlier times. 
  \change{The photons associated with the {\it two} recombinations of helium were released at even higher redshifts. Finding the traces of these signals in the cosmological recombination spectrum will therefore allow us to learn about the state of the Universe at $\sim 130,000$ yrs and $\sim 18,000$ yrs after the big bang. Furthermore, the cosmological recombination radiation may offer a way to tell if something unexpected (e.g. energy release due to annihilating dark matter particles) occurred {\it before} the end of cosmological recombination.}}
\label{RS:fig:plot1}
\end{figure}
\subsection*{Radiation from the cosmological recombination epoch}
\label{RS:sec:Intro4}
In September 1966, one of the authors (RS) was explaining during a seminar at
  the Shternberg Astronomical Institute in Moscow how
  recombination should occur according to the Saha formula for equilibrium ionization.
  After the talk his friend (UV astronomer) {\it Vladimir Kurt} (see Fig.~\ref{RS:fig:ZKS})
  asked him: {\it 'but where are all the redshifted Lyman-$\alpha$ photons
    that were released during recombination?'}
  Indeed this was a great question, which was then addressed in detail
  by \citet{RS_Zeldovich68}, leading to an understanding of the role of the
  2s-two-photon decay, the delay of recombination as compared to the
  Saha-solution (see Fig.~\ref{fig:Xe.Saha} for illustration), 
  the spectral distortions of the CMB due to two-photon
  continuum and Lyman-$\alpha$ emission, the frozen remnant of ionized atoms,
  and the radiation and matter temperature equality until $z\sim 150$.


%
All recombined electrons in hydrogen lead to the release of $\sim
13.6\,$eV in form of photons, but due to the large specific entropy of the Universe
this will only add some fraction of $\Delta \rho_\gamma/\rho_\gamma\sim
10^{-9}-10^{-8}$ to the total energy density of the CMB spectrum, and hence
the corresponding distortions are expected to be very small.
However, all the photons connected with the Lyman-$\alpha$ transition and the
2s-two-photon continuum appear in the Wien part of the CMB spectrum {today},
where the number of photons in the CMB blackbody is dropping exponentially,
and, as realized earlier \citep{RS_Zeldovich68, RS_Peebles68}, these
distortions are significant (see Sect.~\ref{RS:sec:spectrum_all}).

In 1975, {\it Victor Dubrovich} (see Fig.~\ref{RS:fig:VD}) pointed out that
the transitions among highly excited levels in hydrogen are producing
additional photons, which after redshifting are reaching us in the cm- and
dm-spectral band. This band is actually accessible from the ground.
%
Later these early estimates were significantly refined by several groups
(e.g. see \citet{RS_Kholu2005} and \citet{RS_Jose2006} for references), with
the most recent calculation performed by \citet{RS_Chluba2006b}, also
including the previously neglected free-bound component, and showing in detail
that the relative distortions are becoming more significant in the decimeter
Rayleigh-Jeans part of the CMB blackbody spectrum
(see Sects.~\ref{RS:sec:spectrum_all}, Fig.~\ref{RS:fig:DI_results}).
These kind of precise computations are becoming feasible today, because (i)
our knowledge of atomic data (in particular for neutral helium) has
significantly improved; (ii) it is now possible to handle large systems of
strongly coupled differential equations using modern computers; and
(iii) we now know the cosmology model (and most the important parameters like $\Omega_{\rm b}$, $T_\gamma$ and Hubble constant) with sufficiently high precision.
The most interesting aspect of this radiation is that it has a very {\it
  peculiar} but {\it well-defined, quasi-periodic spectral dependence}, where
the photons emitted
due to transitions between levels in the hydrogen atom are coming from redshifts $z\sim 1300-1400$, i.e. {\it before} the
time of the formation of the CMB temperature anisotropies close to the maximum of
the Thomson visibility function (see Fig.~\ref{RS:fig:plot1}).
Therefore, measuring these distortions of the CMB spectrum would provide a way
to confront our understanding of the recombination epoch with {\it direct
  experimental evidence},
and in principle may open another independent way to determine some of the key
parameters of the Universe, \change{like} the value of the CMB monopole
temperature, $T_0$, the number density of baryons, $\propto \Omega_{\rm
  b}h^2$, or alternatively the specific entropy, and the primordial helium
abundance (e.g.  see \citet{RS_Chluba2007d} and references therein).

\begin{figure}
\centering 
\includegraphics[width=0.8\columnwidth]{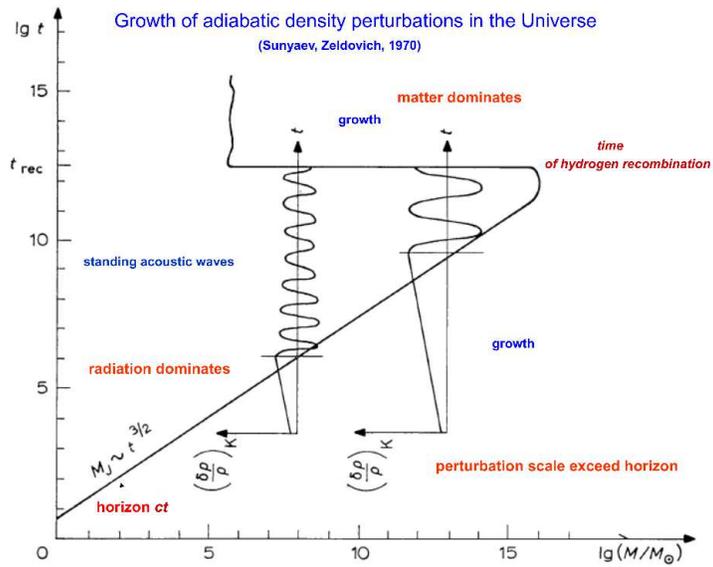}
\\[6mm]
\includegraphics[width=0.8\columnwidth]{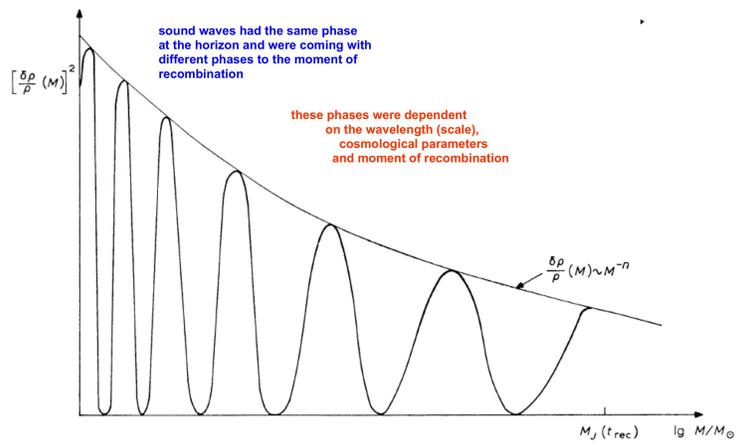}
\caption{Illustration for the growth of adiabatic density perturbations in the Universe. The figure was adapted from \citet{RS_Sunyaev1970}.}
\label{fig:RS.12}
\end{figure}

\begin{figure}
\centering 
\includegraphics[width=0.8\columnwidth]{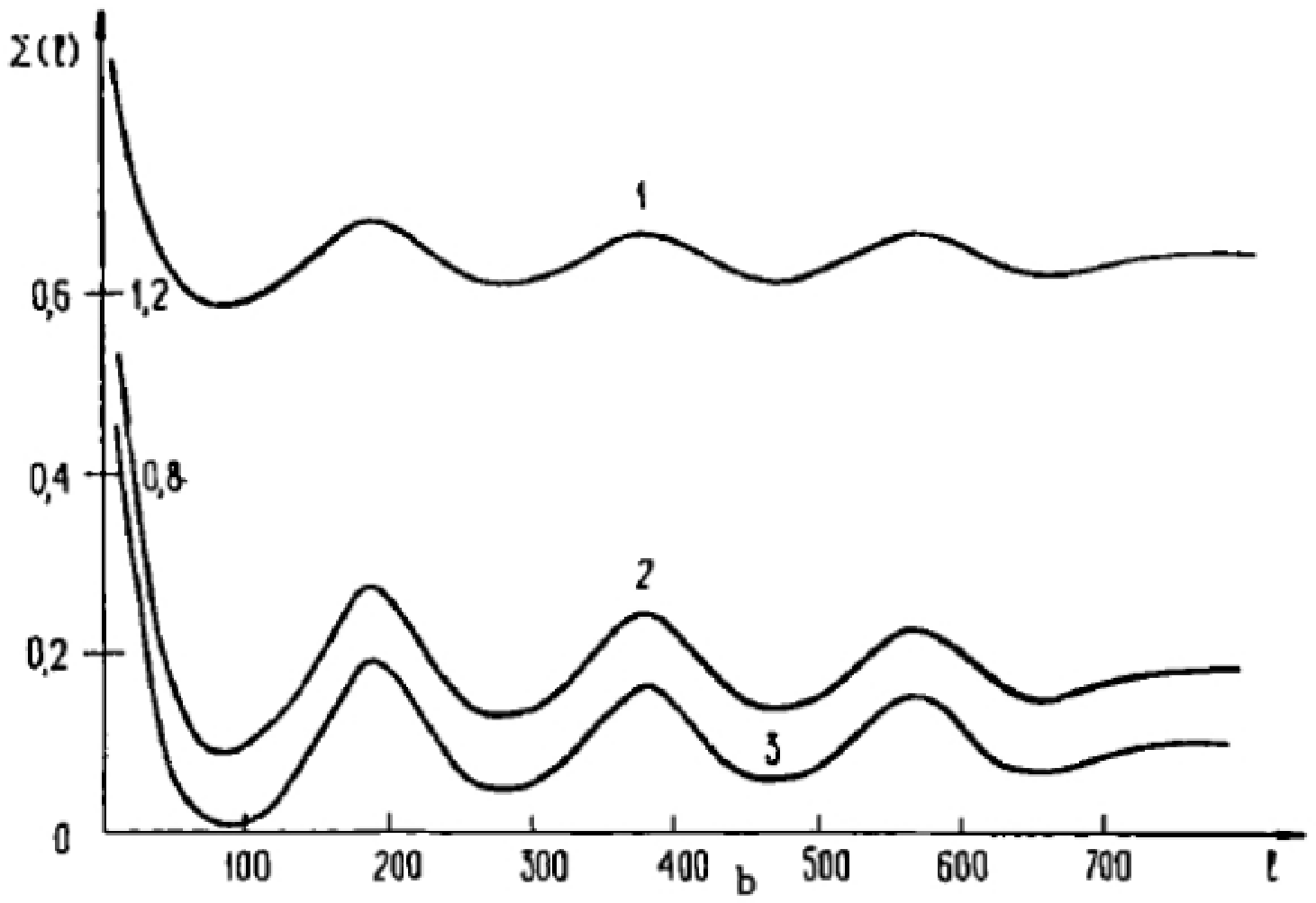}
\\[6mm]
\includegraphics[width=0.8\columnwidth]{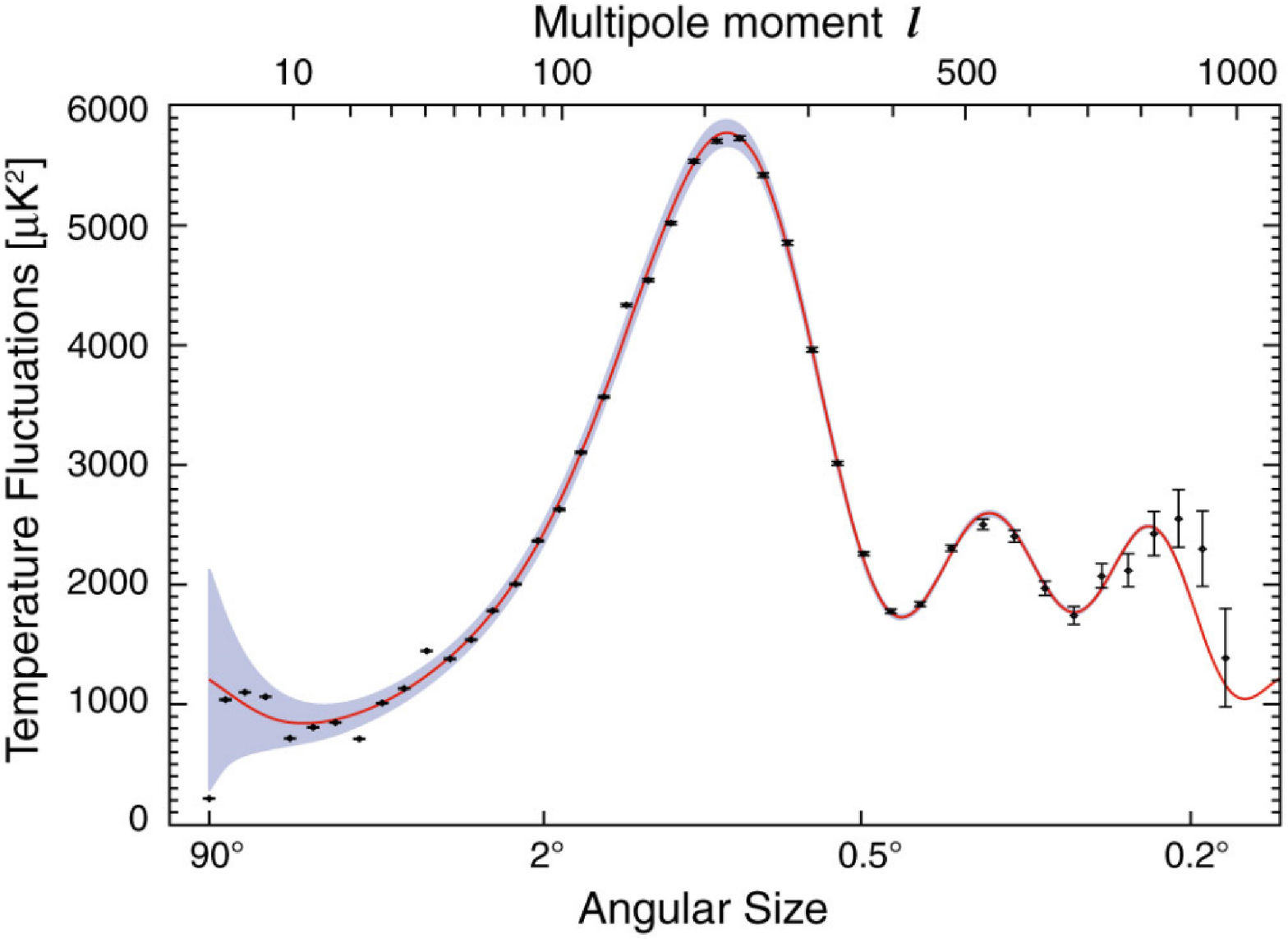}
\caption{First prediction of the acoustic peaks in the spherical harmonics expansion
of the CMB sky map (upper panel) and their modern version with observational data from the WMAP satellite (lower panel). Note that the position of the first peak was already similar to the observed one, although the normalization was completely different.
The figure were taken from \citet{Zeldovich1972} and the WMAP web page 
http://lambda.gsfc.nasa.gov/product/map/current/.}
\label{fig:RS.34}
\end{figure}

\subsection*{Growth of adiabatic density perturbations in the Universe and baryonic acoustic oscillations}
\label{RS:pertub}
It is well known since the classical paper of Eugene Lifshitz in 1946
how adiabatic perturbations are growing in the Universe according to
Einstein's theory of general relativity (GR). 
Nevertheless, it is possible to explain this process using a simple
Newtonian approach and remembering the properties of Jeans gravitational instability. 
At sufficiently early times (but after inflation), when practically any scale of astronomical significance
was bigger than the horizon $ct$ (see upper panel in Fig.~\ref{fig:RS.12}). any two different regions
of the Universe were completely independent. If the densities within them
were different, these independent universes expanded at different rates
and the density differences (perturbations!) were growing according to
a power law.  
The situation changed completely when the perturbation at
a given scale became smaller than the horizon.  
At redshift $z\gtrsim 3300$ our Universe was radiation dominated: 
\change{
the radiation energy density $\epsilon_{\rm r}$ and radiation pressure $\frac{1}{3}\,\epsilon_{\rm r}\approx 0.91 N_\gamma \,k T_\gamma$ significantly exceeded $\epsilon_{\rm b}=\rho_{\rm b}\,c^2$ and the pressure of the
baryons and electrons $\sim 2 N_{\rm b}\,k\,T_\gamma$. 
As mentioned above, the specific entropy of our Universe is huge $N_\gamma/N_{\rm b} \sim 10^9$,  so that 
under these circumstances the {\it sound velocity} 
$v_{\rm s} \sim c/\sqrt{3[1+3\epsilon_{\rm b}/4\epsilon_{\rm r}]}$
was close to speed of light and the {\it Jeans wavelength} was
close to the horizon. 
}
According to the theory of Jeans instability
adiabatic perturbations smaller than Jeans wavelength should \change{evolve} as
{\it sound waves}. GR gives the same answer: the growing mode of
perturbations is initiating standing acoustic waves with wavelengths
depending on the characteristic scale of the perturbation.

These acoustic waves existed till the time of hydrogen recombination. 
After recombination radiation rapidly became free and uniform, Jeans
wavelength, \change{defined by thermal velocities of hydrogen atoms 
$v_{\rm s} \sim \sqrt{2\,kT_\gamma/m_{\rm H}}$, 
decreased many orders of magnitude and only baryons remembered
the {\it phases} which standing acoustic waves had at the moment of
recombination. } 
After recombination density perturbations begun to
grow according to a power law and gave rise to the large scale structure
of the Universe, which we observe today.  
Nevertheless, this characteristic quasi-periodical dependence of the amplitude of
perturbations \change{was conserved up to the phase of nonlinear of growth} (see lower panel in Fig.~\ref{fig:RS.12}).  
This prediction was made by \citet{RS_Sunyaev1970} and in completely
independent way by \citet{RS_Peebles1970}. 
\change{Today we quote this behavior of initial density perturbations as {\it baryonic acoustic oscillations}. It is important to repeat that recombination played crucial role in their appearance.} 

\change{
Simultaneously it was recognized that interaction of CMB photons with with moving electrons and baryon density perturbations must lead to a quasiperiodic dependence of the amplitude of CMB angular fluctuations on angular scale. It was painful for young postdoc (RS) when Zeldovich deleted the words about importance of observational search and added the last phrase into the abstract of the paper by \citet{RS_Sunyaev1970}:  
''{\it A detailed investigation of the spectrum of fluctuations may, in principle, lead to an understanding of the nature of initial density perturbations since a distinct periodic dependence of the spectral density of perturbations on wavelength (mass) is peculiar to adiabatic perturbations. Practical observations are quite difficult due to the smallness of the effects and the presence of fluctuations connected with discrete sources of radio emission}''.
Fortunately Zeldovich told afterwards that physics is beautiful and it is worth to publish this paper. RS was guilty himself because he simultaneously was trying to estimate the angular fluctuations due to presence of radiosources \change{Longair1969}.

A little later, Zeldovich, Rakhmatulina and RS (1972) found that a spherical harmonic
expansion of the future CMB sky map should demonstrate the presence of CMB
acoustic peaks (see upper panel in Fig.~\ref{fig:RS.34}).
Silk damping \citep{RS_Silk1968} was taken into account by \citet{Doroshkevich1978}, who performed realistic computations of acoustic peaks in the baryon dominated Universe.
When it was realized that our Universe contains cold dark matter,
detailed analysis \citep{Peebles1982, Bond1984, Silk1984} showed that baryonic acoustic oscillations should remain important in the modern picture of the Universe.
\change{The predicted acoustic peaks on the CMB sky} were observed in detail  by Boomerang and  MAXIMA1 balloon flights and WMAP spacecraft.  Sloan Digital Sky Survey \citep{Eisenstein2005, Huetsi2006} demonstrated the presence of baryonic oscillations in the spatial distribution of luminous red galaxies.   
}

 \subsection*{The visibility function and its importance}
The Universe was optically thick before recombination, i.e. the mean free path of CMB photons was much smaller than the horizon. After recombination there were practically no free electrons left and the Universe became transparent; since then photons could propagate directly to us.  Hydrogen recombination defines the {\it last scattering surface}.

The  importance of the {\it Thomson visibility function}  $V=\exp(-\tau_{\rm T}) \times {\rm d}\tau_{\rm T}/{\rm d}z$ was recognized already in RS and Zeldovich (1970), when an approximate analytical solution for recombination was found.  This function defines the properties of the {\it last scattering surface}. We should mention here that these beautiful \change{termini} were introduced much later.  
We present the shape of the visibility function in Fig.~\ref{RS:fig:plot1}.

 \subsection*{The era of precision cosmology}
\change{The results from} Boomerang, MAXIMA1 and WMAP together with supernovae 1a observations \citep{Perlmutter1999, Riess1999} and the curve of growth for cluster's of galaxies \citep[see][and references therein]{Vikhlinin2009} opened the {\it era of precision cosmology}, providing detailed information about the key parameters of our Universe.  
It is obvious  that the position and relative amplitude of acoustic peaks is defined by key parameters of the Universe and physical constants. At the same time the corresponding angular separation of acoustic peaks provides us with unique information about the distance to the {\it last scattering surface}.  This demonstrates the great importance of the process of recombination.  Any change in its position or in its sharpness will provide additional and crucial uncertainty in the determination of major parameters of the Universe.  
This is the reason why now we are trying to study the process of recombination with highest possible precision.  
It was a surprise for a majority of theorists that the expected precision of the Planck Surveyor spacecraft will be close to or significantly higher than the precision of widely used present day recombination codes.

\section{The cosmological recombination radiation}
\label{RS:sec:spectrum_all}
%

\begin{figure}
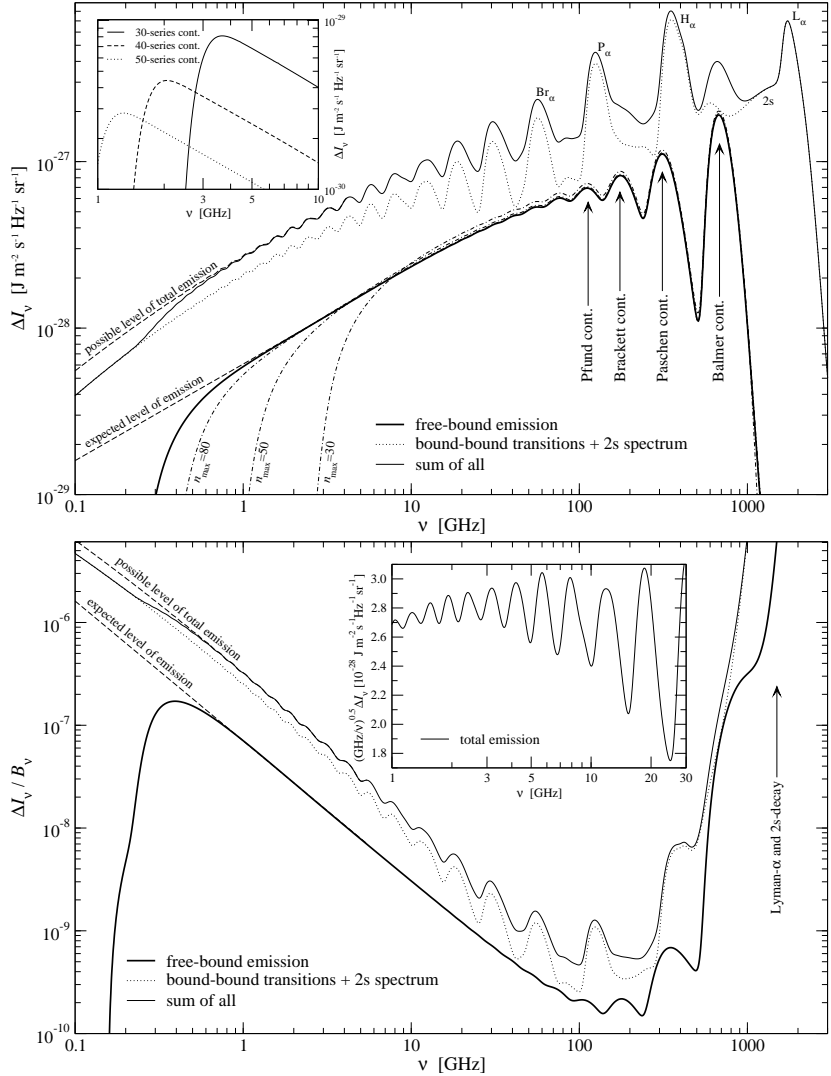

\centering 
\includegraphics[width=0.9\columnwidth]{./RS.DI.eps}
\\
\includegraphics[width=0.9\columnwidth]{./RS.DI_I.eps}
\caption{The full hydrogen recombination spectrum including the free-bound
  emission.
  The results of the computation for 100 shells were used.
The contribution due to the 2s two-photon decay is also accounted for.
The dashed lines indicate the expected level of emission when including more
shells. In the upper panel we also show the free-bound continuum spectrum for
different values of $n_{\rm max}$ (dashed-dotted). The inlay gives the
free-bound emission for $n=30,\,40$, and $50$.
The lower panel shows the distortion relative to the CMB blackbody spectrum,
and the inlay illustrates the modulation of the total emission spectrum for
$1\,\text{GHz}\leq \nu \leq 30\,\text{GHz}$ in convenient coordinates. The
figure is from \citet{RS_Chluba2006b}.}
\label{RS:fig:DI_results}
\end{figure}
\subsection{Contributions due to standard hydrogen recombination}
\label{RS:sec:spectrum}
Within the picture described above it is possible to compute the {\it
  cosmological hydrogen recombination spectrum} with high accuracy.
  \change{The photons corresponding to this spectral distortion of the CMB have been emitted mostly at redshifts $z\sim 1300-1400$, and therefore reach the observer today $\sim 10^3$ times redshifted.}
In Figure \ref{RS:fig:DI_results} we give the results of our computations for
frequencies from $100\,$MHz up to $3000\,$GHz.
The free-bound and bound-bound atomic {\it transitions among 5050 atomic
  levels} had to be taken into account in these computations.
At high frequencies one can clearly see the features connected with the
Lyman-$\alpha$ line, and the Balmer-, Paschen- and Brackett-series, whereas
below $\nu\sim 1\,$GHz the lines coming from transitions between highly
excited level start to merge to a continuum.
Also the features due to the Balmer and the 2s-1s two-photon continuum are
visible.
\change{
Overall the free-bound emission contributes about 20\%-30\% to the spectral
distortion due to hydrogen recombination at each frequency, and a total of
$\sim 5$ photons per hydrogen atom are released in the full hydrogen
recombination spectrum.
}

\begin{figure}
\centering 
\includegraphics[width=0.80\columnwidth]{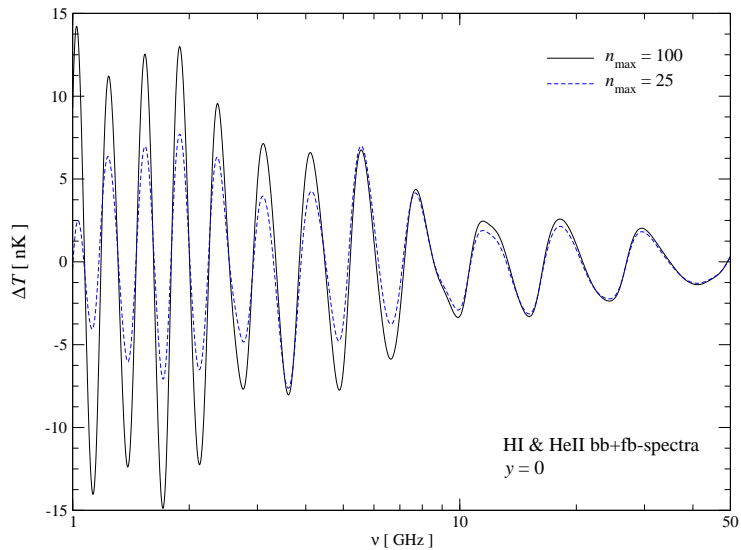}
\caption{Frequency-dependent modulation of the CMB temperature caused by photons from the \ion{H}{i} and \ion{He}{ii} recombination epochs. Both the bound-bound and free-bound contributions were included, and the mean recombination spectrum was subtracted. The shown signal is practically {\it unpolarized} and the same in {\it all} directions on the sky. The figure is taken from \citet{Chluba2008a}.}
\label{fig:DT_n_100_25}
\end{figure}
%
One can also see from Figure \ref{RS:fig:DI_results} that both in the Wien and
the Rayleigh-Jeans region of the CMB blackbody spectrum the relative
distortion is growing. In the vicinity of the Lyman-$\alpha$ line the relative
distortion exceeds unity by several orders of magnitude, but unfortunately at
these frequencies the cosmic infra-red background due to sub-millimeter, dusty
galaxies renders a direct measurement impossible.
Similarly, around the maximum of the CMB blackbody at $\sim 150\,$GHz it will
\change{likely} be hard to measure these distortions with current technology, although there
the spectral variability of the recombination radiation is largest.  
However, at low frequencies ($\nu\lesssim 2\,$GHz) the relative distortion exceeds the level of
$\Delta I/I\sim 10^{-7}$ but still has variability with well-defined frequency dependence at a level of several
percent.

\change{As additional example, the total recombination spectrum from hydrogen and \ion{He}{ii} at frequencies in the range $1\,\text{GHz}\lesssim \nu \lesssim 10\,$GHz leads to a frequency-dependent modulation of the CMB temperature by $\Delta T\sim \pm 5-15\,$nK (see Fig.~\ref{fig:DT_n_100_25} for more details), where the signal is expected to have many spectral features over one octave or one decade in frequency. These signatures from the cosmological recombination epochs are very hard to mimic by other astrophysical sources or instrumental noise, so that it may become possible to extract then in the future (see Sect.~\ref{RS:sec:spectrum_strat} for illustration of a possible observing strategy).}

{
\begin{figure}
\centering 
\includegraphics[height=\columnwidth, angle=90]{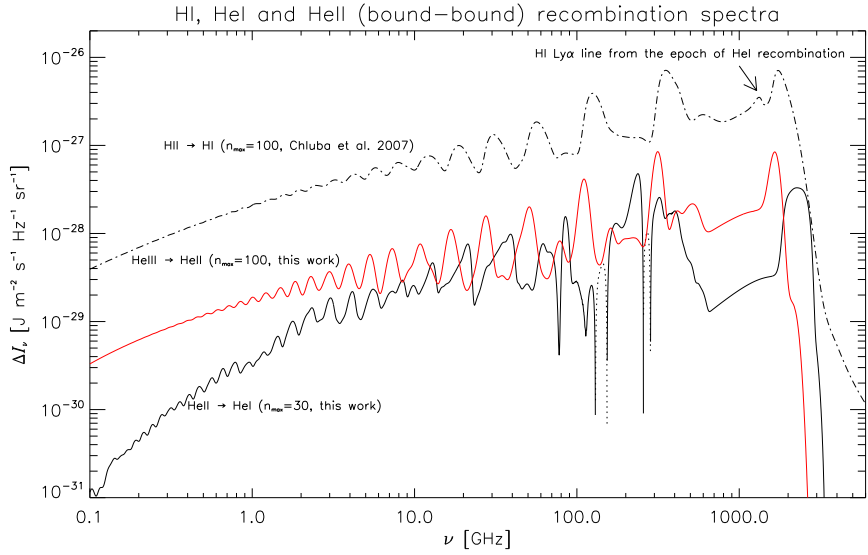}
\caption{Helium and hydrogen (bound-bound) recombination spectra. The
  following cases are shown: (a) the $\ion{He}{ii}\rightarrow \ion{He}{i}$
  recombination spectrum (black solid line), which has been obtained including
  up to $n_{\rm max}=30$ shells, and considering all the J-resolved
  transitions up to $n=10$. In this case, there are two negative features,
  which are shown (in absolute value) as dotted lines; (b) the
  $\ion{He}{iii}\rightarrow\ion{He}{ii}$ recombination spectrum (red solid
  line), where we include $n_{\rm max}=100$ shells, resolving all the angular
  momentum sub-levels and including the effect of Doppler broadening due to
  scattering off free electrons; (c) the \ion{H}{i} recombination spectrum,
  where we plot the result from \citet{RS_Chluba2007} up to $n_{\rm max}=100$.
  The \ion{H}{i} Lyman-$\alpha$ line arising in the epoch of \ion{He}{i}
  recombination is also added to the hydrogen spectrum (see the feature around
  $\nu=1300$~GHz). In all three cases, the two-photon decay continuum of the
  $n=2$ shell was also incorporated. The figure is taken from
  \citet{RS_Jose2007}.}
\label{RS:fig:HeSpec}
\end{figure}
\begin{figure}
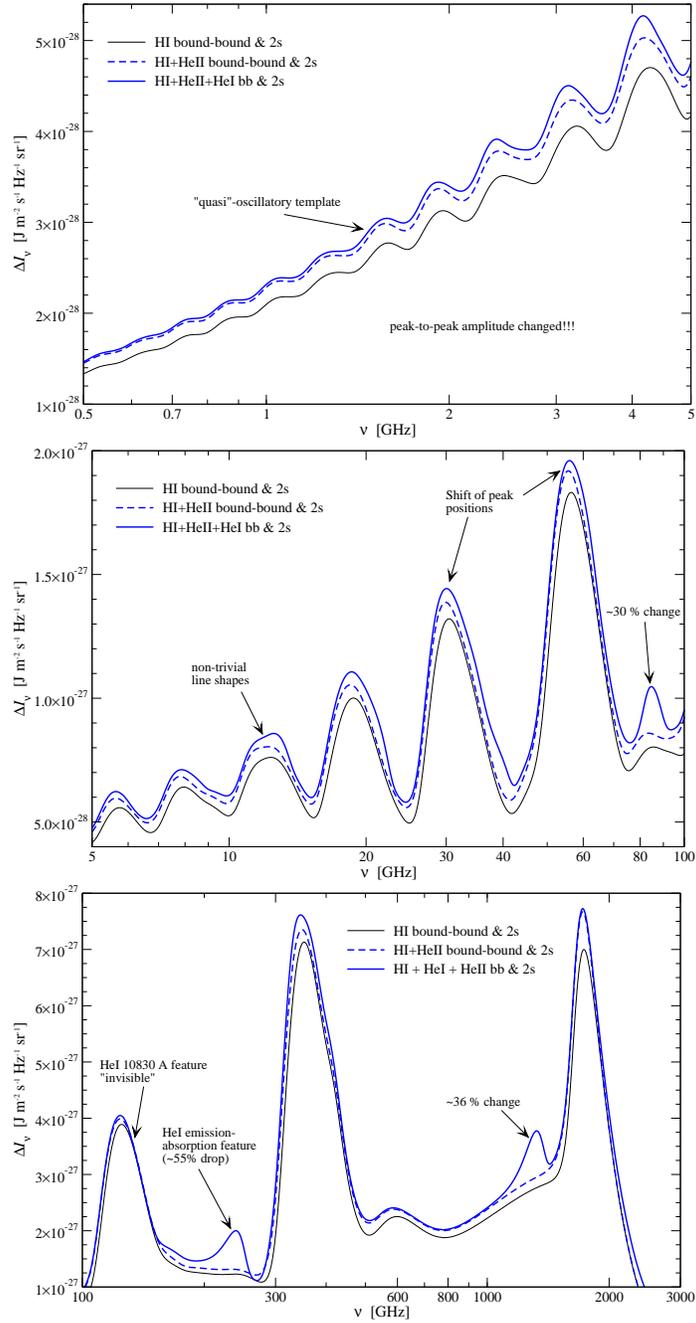

\centering 
\includegraphics[width=0.75\columnwidth]{./DI.total.low.eps}
\\
\includegraphics[width=0.75\columnwidth]{./DI.total.high.eps}
\\
\includegraphics[width=0.75\columnwidth]{./DI.total.very_high.eps}
\caption{Helium and hydrogen (bound-bound) recombination spectra in different
frequency bands. The curves where obtained summing the results shown in
Fig.~\ref{RS:fig:HeSpec}. In the figures we also pointed out some of the most
significant additions to the pure hydrogen recombination spectrum, which are
only because of the presence of pre-stellar helium in the primordial plasma.}
\label{RS:fig:HeSpeczooms}
\end{figure}

\subsection{The contributions due to standard helium recombination}
\label{RS:sec:He_spectrum}
Why would one expect any significant contribution to the cosmological
recombination signal from helium, since it adds only $\sim 8\%$ to the
total number of atomic nuclei?
First of all, there are {\it two} epochs of helium recombination, i.e. $1600
\lesssim z\lesssim 3500$ for \ion{He}{ii}$\rightarrow$\ion{He}{i} and $5000
\lesssim z\lesssim 8000$ for \ion{He}{iii}$\rightarrow$\ion{He}{ii}
recombination. 
Therefore, overall one can already expect some $\sim 16\%$ contribution to the
recombination spectrum due to the presence of helium in the Universe.
However, it turns out that in some spectral bands the total emission due to
helium transitions can reach amplitudes up to $\sim 30\%-50\%$
\citep{RS_Jose2007}. This is possible, since
\ion{He}{iii}$\rightarrow$\ion{He}{ii} actually occurs much faster, following
the Saha-solution much closer than in the case of hydrogen recombination. 
\change{Therefore photons
are emitted in a narrower range of frequencies, and even the line broadening due
to electron scattering cannot alter the shape of the features significantly until today (see
Fig.~\ref{RS:fig:HeSpec}).}

In addition, the recombination of neutral helium is sped up due to the
absorption of $2^1{\rm P}_1-1^1{\rm S}_0$ and $2^3{\rm P}_1-1^1{\rm
S}_0$-photons by the tiny fraction of neutral hydrogen already present at
redshifts $z\lesssim 2400$. This process was suggested by P.~J.~E. Peebles in
the mid 90's \citep[see remark in][]{Hu1995}, but only recently \change{it has been} convincingly
taken into account by \citet{RS_HirataI} and others \citep{RS_Kholu2007,
RS_Jose2007}.
This also makes the neutral helium lines more narrow and enhances the emission
in some frequency bands (see Fig.~\ref{RS:fig:HeSpec} and for more details Fig.~\ref{RS:fig:HeSpeczooms}).
Also the re-processing of \change{helium photons} by hydrogen lead to additional signatures
in the recombination spectrum, most prominently the 'pre-recombinational'
\ion{H}{i} Lyman-$\alpha$ line close to $\nu\sim1300\,$GHz (see
Fig.~\ref{RS:fig:HeSpeczooms}).

\change{We would like to mention, that} the first computations of the helium recombination spectrum were performed by
\citet{RS_Dubrovich1997}, before the cosmological concordance model was
\change{actually} established. Also neutral helium recombination was still considered to occurs
much slower, since the effect connected to the hydrogen continuum opacity was
not taken into account, and the existing atomic data for \ion{He}{i} was still
rather poor.
In the most recent computations of the neutral helium spectrum
\citep{RS_Jose2007}, for both the singlet and triplet atom, up to $n_{\rm
max}=30$ shells were included. This amounts in a total of $\sim 1000$
different atomic levels. Furthermore, we have taken into account all
fine-structure and \change{most of the} singlet-triplet transitions for levels with $n\leq 10$,
using the atomic data published by \citet{Drake2007} and according to the
approach discussed with \citet{RS_BeigVain}.
In the case of neutral helium, the non-trivial superposition of all lines even
lead to the appearance of two {\it negative features} in the total \ion{He}{i}
bound-bound recombination spectrum (see Fig.~\ref{RS:fig:HeSpec}). The one at
$\nu\sim 145\,$GHz is coming from one of the $10830~\AA$ fine-structure lines,
whereas the feature close to $\nu\sim 270\,$GHz is mainly due to the
superposition of the negative $5877~\AA$ and positive $6680~\AA$-lines
\citep{RS_Jose2007}.
}

\begin{figure}
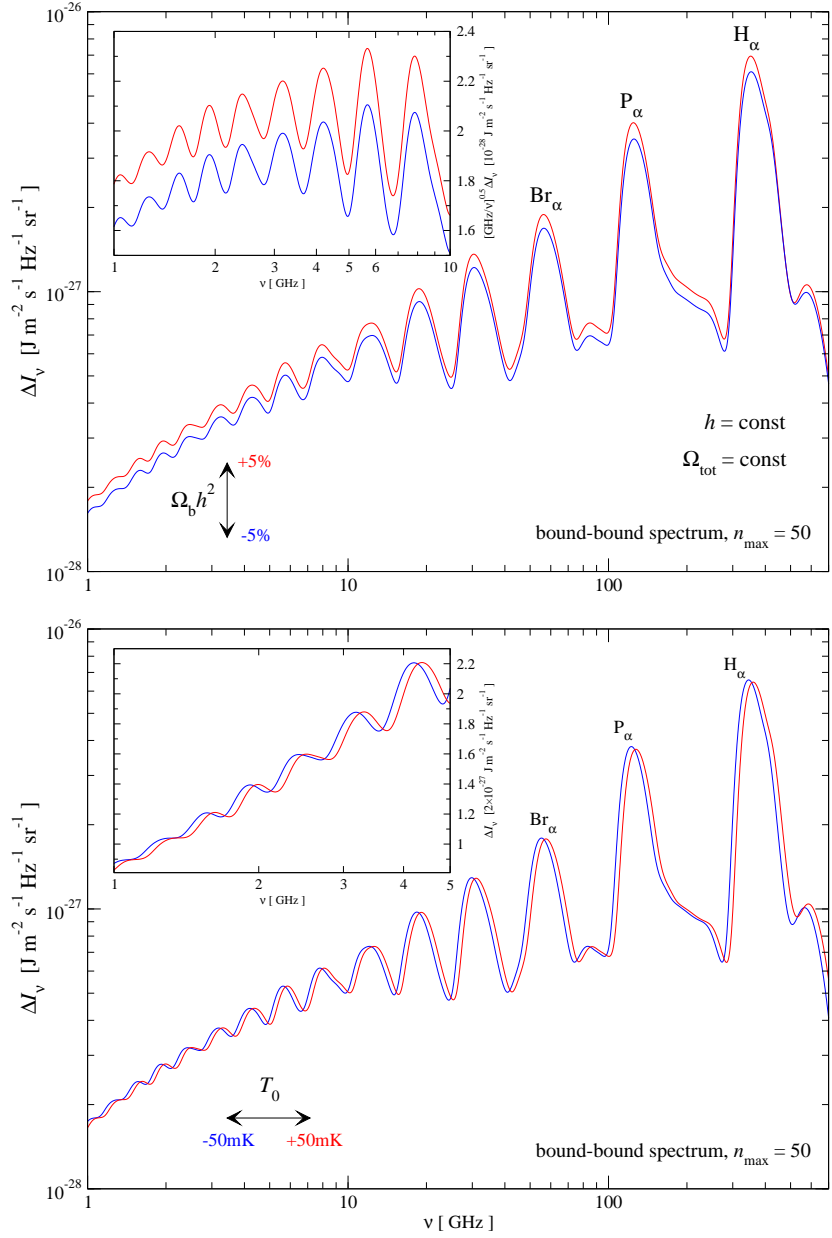

\centering 
\includegraphics[width=0.9\columnwidth]{./RS.DJ.Ob.50.eps}
\\
\includegraphics[width=0.9\columnwidth]{./RS.DJ.T0.50mK.50.eps}
\caption{The bound-bound hydrogen recombination spectrum for $n_{\rm
    max}=50$. The upper panel illustrates the dependence on $\Omega_{\rm
    b}h^2$, and the lower the dependence on the value of $T_0$. 
The figure is from \citet{RS_Chluba2007d}.}
\label{RS:fig:DI_cosmos}
\end{figure}
\subsection{Dependence of the recombination spectrum on cosmological parameters}
\label{RS:sec:spectrum_cosmos}
In this Section we want to {\it illustrate} the impact of different
cosmological parameters on the hydrogen recombination spectrum. We \change{restricted}
ourselves to the bound-bound emission spectrum and included 50 shells for
the hydrogen atom into our computations.

In Fig.~\ref{RS:fig:DI_cosmos} we illustrate the dependence of the hydrogen
recombination spectrum on the value of the CMB monopole temperature, $T_0$.
The value of $T_0$ mainly defines the time of recombination, and consequently
when most of the emission in each transition appears. This leads to a
dependence of the {\it line positions} on $T_0$, but the total intensity in each
transition (especially at frequencies $\nu \lesssim 30\,$GHz) remains
practically the same. We found that the fractional shift of the low frequency
spectral features along the frequency axis scales roughly like $\Delta
\nu/\nu\sim\Delta T/T_0$. 
\change{Hence $\Delta T\sim1\,$mK implies $\Delta\nu/\nu\sim 0.04\,\%$ or
$\Delta\nu\sim 1\,$MHz at $2\,$GHz, which with modern spectrometers is rather easy to resolve.}
Since the maxima and minima of the line features due to the large duration of
recombination are rather broad ($\sim 10\%-20\%$), it is probably better to look
for these shifts close to the steep parts of the lines, where the derivatives
of the spectral distortion due to hydrogen recombination are largest.
It is also important to mention that the hydrogen recombination spectrum is
shifted as a {\it whole}, allowing to increase the significance of a
measurement by considering many spectral features at several frequencies.

We showed in \citet{RS_Chluba2007d} that the cosmological hydrogen
recombination spectrum is practically independent of the value of $h$. Only
the features due to the Lyman, Balmer, Paschen and Brackett series are
slightly modified.
This is connected to the fact, that $h$ affects the ratio of the atomic
time-scales to the expansion time. Therefore changing $h$ affects the escape
rate of photons in the Lyman-$\alpha$ transition and the relative importance
of the 2s-1s transition. For transitions among highly excited states it is not
crucial via which channel the electrons finally reach the ground state of
hydrogen and hence the modifications of the recombination spectrum at low
frequencies due to changes of $h$ are small.
Changes of $\Omega_{\rm m}h^2$ should affect the recombination spectrum for
the same reason.

The lower panel in Fig.~\ref{RS:fig:DI_cosmos} illustrates the dependence of
the hydrogen recombination spectrum on $\Omega_{\rm b}h^2$. It was shown that
the total number of photons released during hydrogen recombination is directly
related to the total number of hydrogen nuclei \citep[e.g.][]{RS_Chluba2007d}.
Therefore one expects that the overall {\it normalization} of the recombination
spectrum depends on the total number of baryons, $N_{\rm b}\propto\Omega_{\rm
  b}h^2$, and the helium to hydrogen abundance ratio, $Y_{\rm p}$.
Varying $\Omega_{\rm b}h^2$ indeed leads to a change in the overall amplitude
$\propto \Delta(\Omega_{\rm b}h^2)/(\Omega_{\rm b}h^2)$.
Similarly, changes of $Y_{\rm p}$ should affect the normalization of the
hydrogen recombination spectrum, but here it is important to also take the
helium recombination spectrum into account. 
\change{
Like in the case of hydrogen there is an effective number of photons
that is produced per helium atom during \ion{He}{iii}$\rightarrow$\ion{He}{ii}
and \ion{He}{ii}$\rightarrow$\ion{He}{i} recombination. Changing $Y_{\rm p}$
will affect the relative contribution of hydrogen and helium to the cosmological
recombination spectrum. Since the physics of helium recombination is different
than in the case of hydrogen (e.g. the spectrum of neutral helium is more
complicated; helium recombination occurs at earlier times, when the medium was
hotter; \ion{He}{iii}$\rightarrow$\ion{He}{ii} is more rapid, so that the
recombination lines are more narrow), one can expect to
find direct evidence of the presence of helium in the full recombination
spectrum. These might be used to quantify the total amount of helium during
the epoch of recombination, well before the first appearance of stars.
}

\begin{figure}
\centering 
\includegraphics[width=0.75\columnwidth]{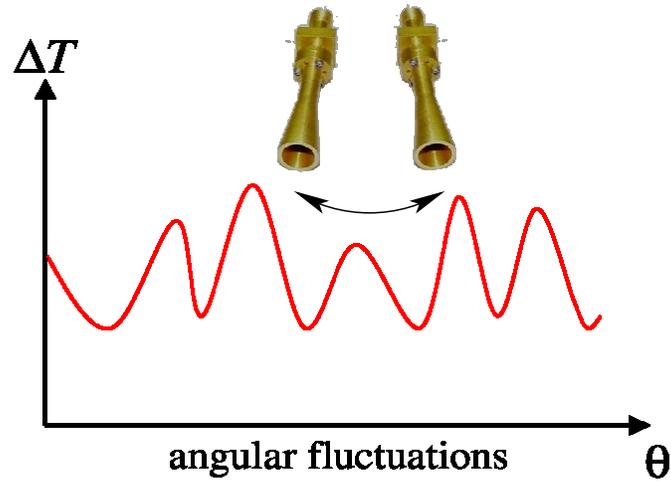}
\\[8mm]
\includegraphics[width=0.75\columnwidth]{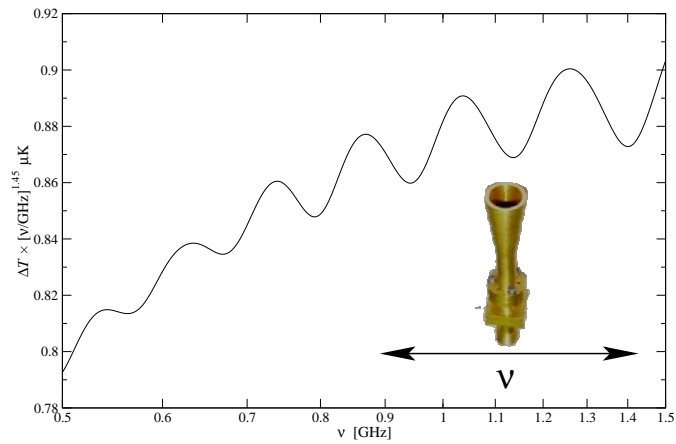}
\caption{Comparison of observing strategies: top panel -- observations of the
  CMB temperature anisotropies. Here one is scanning the sky at fixed frequency in
  different directions. lower panel -- proposed strategy for the signal from
  cosmological recombination. For this one may fix the observing direction,
  choosing a large, least contaminated part of the sky, and scan along the
  frequency axis instead.}
\label{RS:fig:strategy}
\end{figure}
\subsection{A possible observing strategy}
\label{RS:sec:spectrum_strat}
In order to measure {the distortions under discussion} one should scan
the CMB spectrum along the {\it frequency axis} including several spectral
bands (for illustration see Fig.~\ref{RS:fig:strategy}). Because the CMB
spectrum is the {\it same in all directions}, one can collect the flux of large
regions on the sky, particularly choosing patches that are the least
contaminated by other astrophysical foregrounds.
\change{Also the recombinational signals should be practically {\it unpolarized}, a fact that provides another way to distinguish it from other possible contaminants.}
{\it No absolute measurement} is necessary, but one only has to look for a
modulated signal at the $\sim\mu$K level, with typical \change{peak-to-peak} amplitude of
$\sim 10-30\,$nK and $\Delta \nu/\nu\sim 0.1$ (e.g. see Fig.~\ref{fig:DT_n_100_25}), where this signal can be predicted
with high accuracy, yielding a {\it spectral template} for the full
cosmological recombination spectrum, which should also include the
contributions from helium.
\change{Note that} for observations of the CMB temperature anisotropies a sensitivity level of
$10\,$nK in principle can be already achieved \citep{RS_ReadheadPC}.

\change{
We want to stress again, that measuring these distortions of the CMB
spectrum would provide a way to confront our understanding of the
recombination epoch with {\it direct experimental evidence},
and in principle may deliver another independent method to determine some of the key
parameters of the Universe, in particular the value of the CMB monopole
temperature, $T_0$, the number density of baryons, $\propto \Omega_{\rm
  b}h^2$, and the pre-stellar helium
abundance, {\it not suffering} from limitations set by {\it cosmic
    variance} (see Sect.~\ref{RS:sec:spectrum_cosmos} for more details).
As we will explain in the next section, most importantly if something {\it non-standard} occurred during or before the epoch of cosmological recombination, this should leave some potentially observable traces in the cosmological recombination radiation, which would allow us to learn additional details about the {\it thermal history} of our Universe.
}

\begin{figure}
\centering 
\includegraphics[width=0.75\columnwidth]{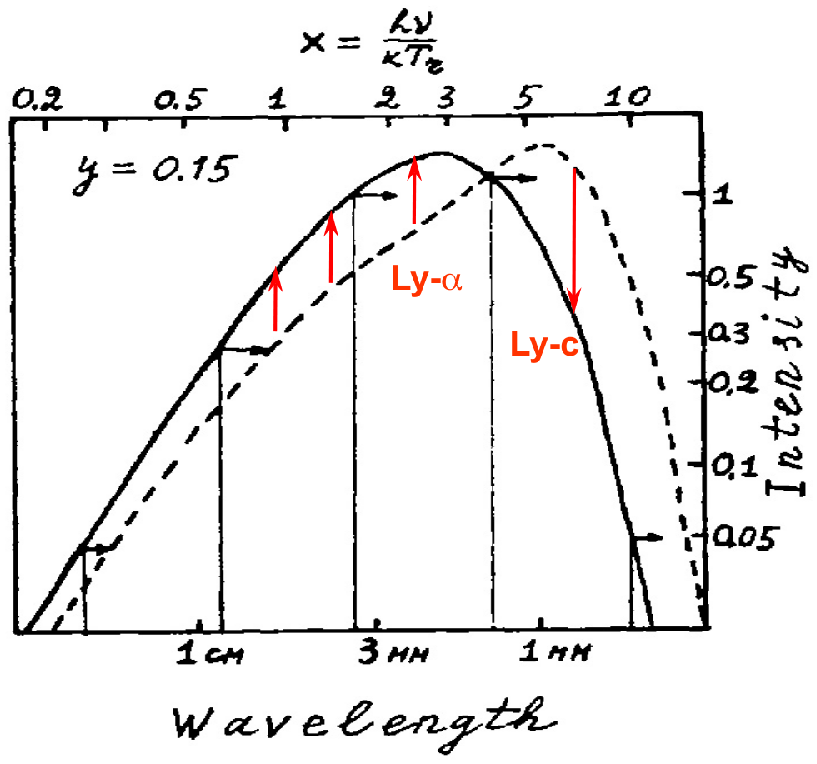}
\caption{Illustration of a Compton $y$-distortion for $y=0.15$. The solid line shows that CMB blackbody spectrum, while the dashed line represents the distorted CMB spectrum.
The figure was adapted from \citet{Sunyaev1980}.
}
\label{fig:ydist}
\end{figure}

\begin{figure}
\centering 
\includegraphics[width=0.44\columnwidth]{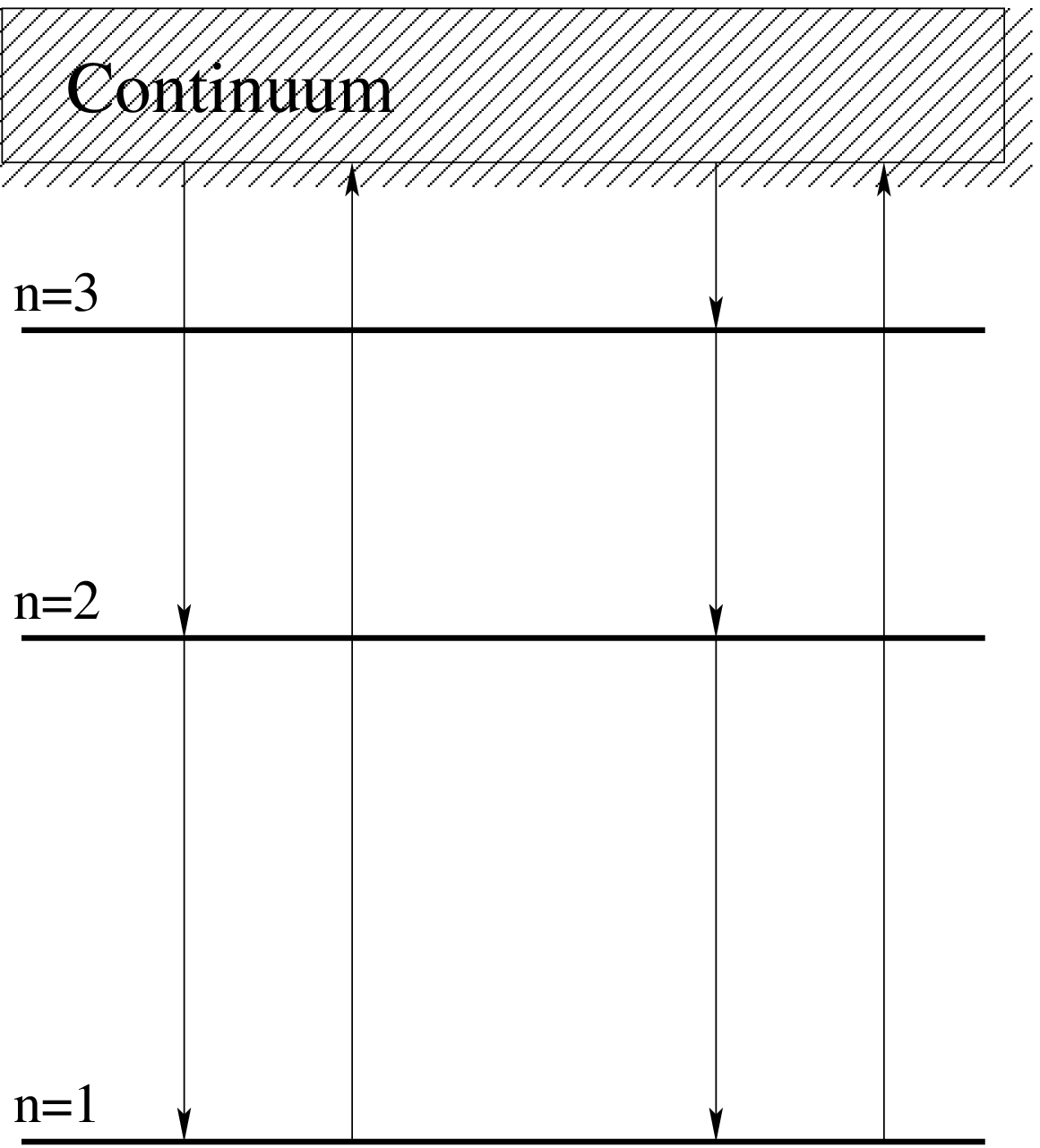}
\hspace{3mm}
\includegraphics[width=0.44\columnwidth]{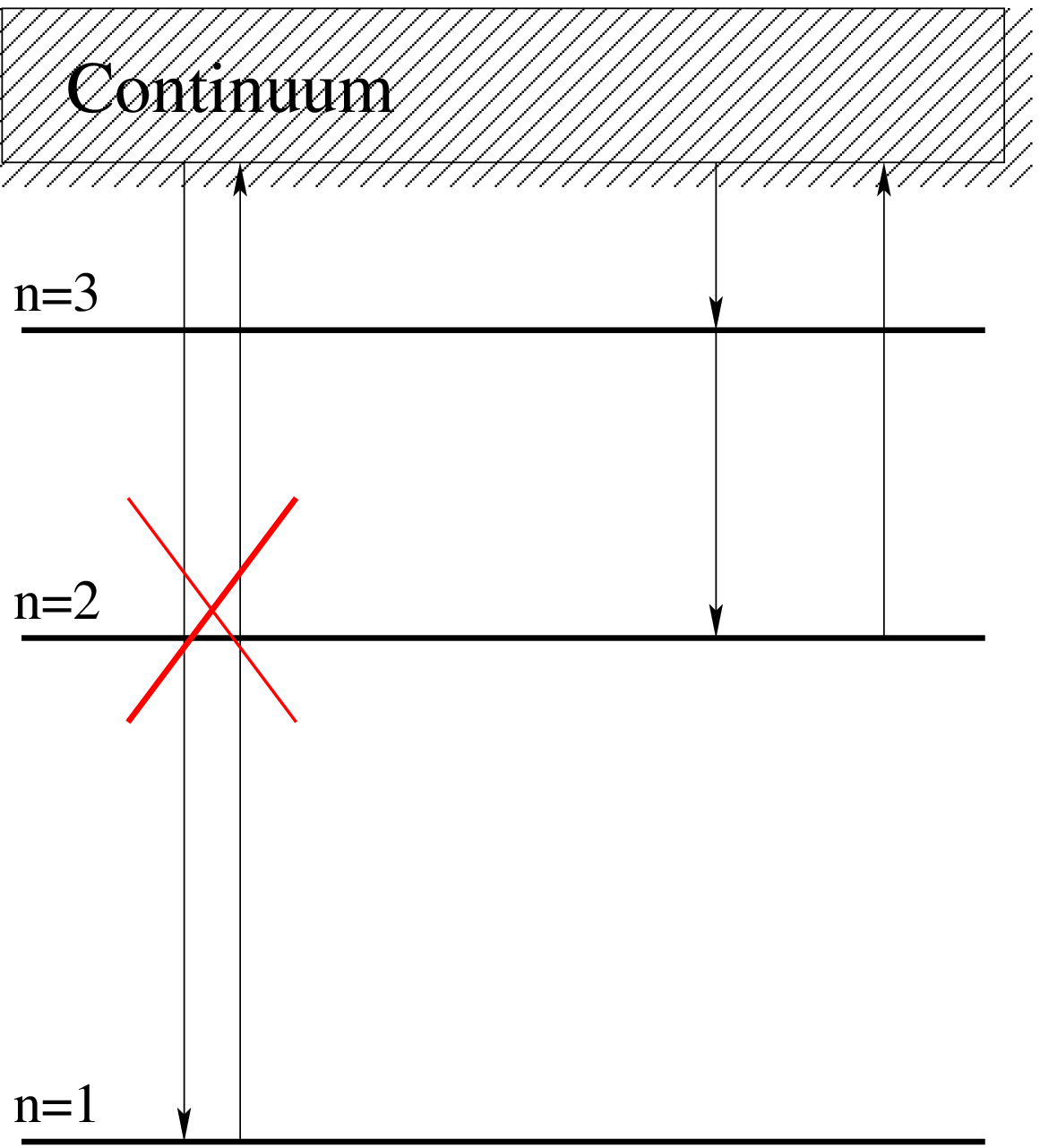}
\caption{Sketch of the main atomic loops for hydrogen and \ion{He}{ii} when
  including 3 shells. The left panel shows the loops for transitions that are
  terminating in the Lyman-continuum. The right panel shows the case, when the
  Lyman-continuum is completely blocked, and unbalanced transitions are
  terminating in the Balmer-continuum instead.
  In the first case up to 3 photons can be created per absorbed Lyman continuum photon, while in the later 2 photons are released per absorbed Balmer continuum photon.
  \change{The figure was taken from \citet{Chluba2008a}}
  }
\label{fig:loops}
\end{figure}
\subsection{The cosmological recombination radiation after energy release before the end of hydrogen recombination}
\label{RS:sec:prespectrum}
All the computations for the standard cosmological recombination spectrum presented in the previous sections were performed assuming that at all times the ambient CMB radiation field is given by a {\it pure blackbody} spectrum with temperature $T_\gamma\propto (1+z)$. Furthermore, it is assumed that the distortions created in the recombination epochs are negligibly small, except for those from the main resonances, e.g. the Lyman series in the case of hydrogen. 
These assumptions are very well justified for a {\it standard thermal history} of the Universe, since the expansion of the Universe alone does not alter the shape of the photon distribution.
Therefore it is clear that  well {\it before} the recombination epoch atomic emission and absorption processes are balancing each other with extremely high precision, so that no net signal to the CMB spectrum can be created.

However, it is well known that the CMB spectrum in principle could deviate from a pure blackbody, if at some point some {\it energy release} (e.g. due to decaying or annihilating particles) occurred, leading to a {\it non-standard thermal history} of the Universe.
For early energy release ($\pot{5}{4}\lesssim z \lesssim
\pot{2}{6}$) the resulting spectral distortion can be characterized as a
Bose-Einstein $\mu$-type distortion \citep{Sunyaev1970b, Illarionov1975a,
  Illarionov1975b}, while for energy release at low redshifts ($z\lesssim
\pot{5}{4}$) the distortion is close to a $y$-type distortion
\citep{Zeldovich1969}.
The current best observational limits on these types of distortions were obtained using the {\sc Cobe/Firas} instrument, yielding $|y|\leq \pot{1.5}{-5}$ and $|\mu|\leq \pot{9.0}{-5}$ \citep{Fixsen1996}.
Here we now want to address the question of how a $y$-distortion with $y\lesssim \pot{1.5}{-5}$ would affect the cosmological recombination radiation and what one could learn about the mechanism that lead to the energy injection by observing the recombinational radiation.

\subsubsection*{Transition loops in a non-blackbody ambient radiation field}
\label{RS:sec:loops}
If we assume that at redshift $z_{\rm i}\lesssim \pot{5}{4}$ some amount of energy was released, then afterwards the intrinsic CMB spectrum deviates from a pure blackbody, where the spectral distortion will be given by a $y$-type distortion.
The $y$-parameter will be directly related to the total amount of energy that was released, but here it only matters that it does not exceed the upper limit given by {\sc Cobe/Firas}.
In comparison to the blackbody spectrum a $y$-type distortion\footnote{
This type of CMB distortion is also well known in connection with the thermal SZ-effect caused by the scattering of CMB photons by the hot electron plasma inside the deep potential wells of clusters of galaxies \citep{Sunyaev1980}.
}
is characterized by a {\it deficit} of photons at low and an {\it increment} at high frequencies (see Fig.~\ref{fig:ydist} for illustration).

It is clear that {\it after} the energy release the equilibrium between the matter and radiation is perturbed, and a small imbalance between atomic emission and absorption is created, which leads to the development of closed {\it loops} of transitions \citep{Liubarskii83}.
These loops can now produce a {\it net change} in the number of photons, even {\it prior} to the epoch of recombination, but otherwise they leave the ionization degree of the Universe unaltered.
Also, it is expected that they should always form in such a way that the net destruction and creation of photons will tend to re-establish the full equilibrium between matter and radiation. 
As an example, if we consider a redshift at which the Lyman continuum frequency of hydrogen is located in the Wien part of the distorted CMB, while the other transitions are still in the Rayleigh-Jeans part of the spectrum (see Fig.~\ref{fig:ydist} for illustration), then the excess of Lyman continuum photons over the value for the blackbody, will lead to an excess photo-ionization of hydrogen atoms from the ground-state.
On the other hand, the deficit of photons in the low frequency part of the background radiation spectrum will allow slightly more electrons to be captured to (highly) excited states than for a blackbody ambient radiation field. 
From these excited states the electrons then can cascade down towards the ground state, emitting several low frequency photons during the dipole transition via intermediate levels.
In this closed loop which started with the destruction of  a Lyman-continuum photon, several low frequency photons can be created (see Fig.~\ref{fig:loops}).

\subsubsection*{What one could learn from the pre-recombinational recombination radiation}
The described process is expected to alter the total radiation coming from atomic transitions in the early Universe and may leave some observable spectral features in addition to those produced during the normal recombination epoch (see next section for details). 
The interesting point is that the photons which are created in these loops are emitted in the {\it pre-recombinational epoch} of the considered atomic species. Therefore, it will make a difference, if energy injection occurred {\it before} $\ion{He}{iii}\rightarrow\ion{He}{ii}$ recombination, at different stages {\it between} the three recombination epochs, or {\it after} hydrogen recombination finished (see Fig.~\ref{RS:fig:plot1} for reminder on the different recombination epochs).
In particular, if energy injection occurred after hydrogen recombination finished, then there should be {\it no additional} traces of this energy injection in the recombinational radiation. 
This fact provides the interesting possibility to distinguish a pre-recombinational all sky $y$-distortion from the one that is created e.g. due to unresolved SZ-clusters, supernova explosions, or the warm-hot-intergalactic medium at redshift well below the recombination epoch.
Here it is important that a normal $y$-distortion is {\it completely featureless}, so that it is very hard to tell when the distortion was introduced. 
However, the changes in the cosmological recombination radiation generated by energy injection before the end of hydrogen recombination not only depend on the {\it amount of energy} that was injected but also on the {\it time} and {\it duration} of this process \citep{Chluba2008a}.
Furthermore, energy injection does leave {\it distinct spectral features} in the recombinational radiation (see next section), which may allow us to learn much more than just confirming that there was some energy injection at some point.

\begin{figure}
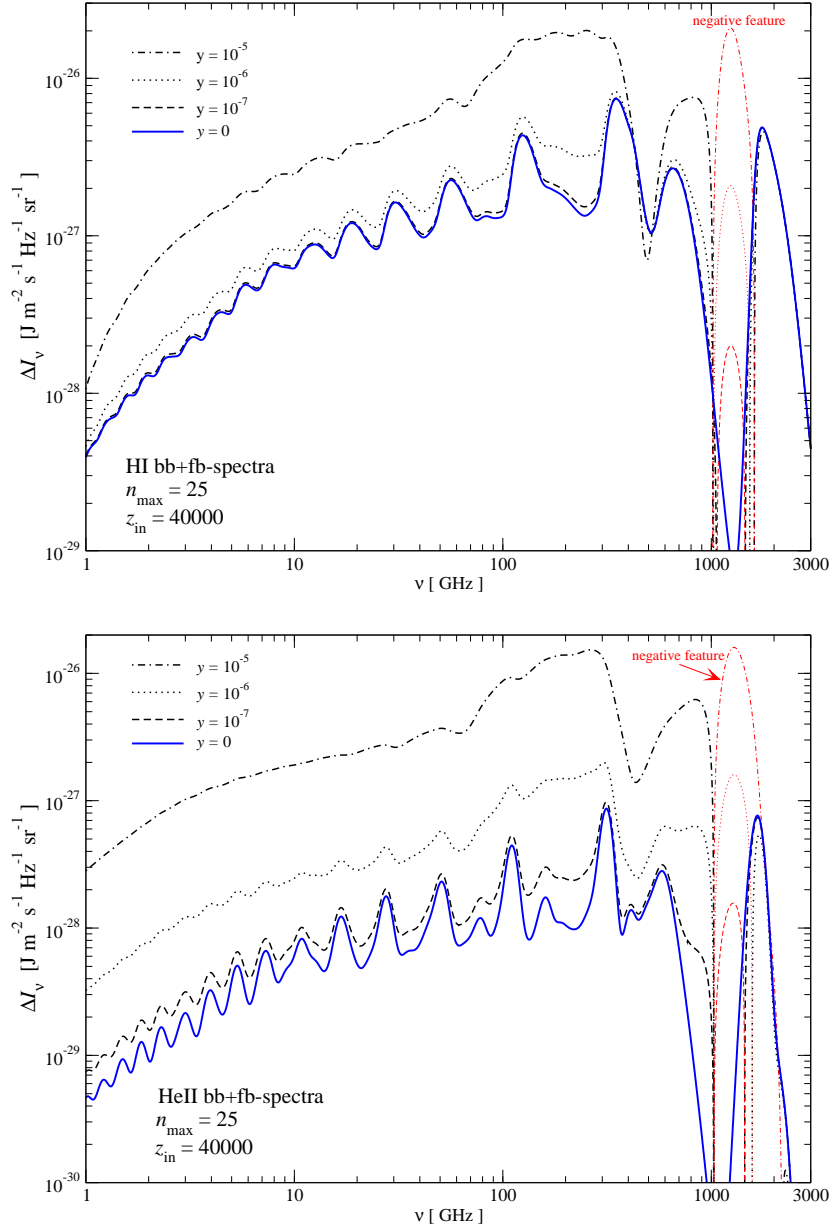

\centering 
\includegraphics[width=0.9\columnwidth]{./spec.25.diff_y.HI.sum.eps}
\\[5mm]
\includegraphics[width=0.9\columnwidth]{./spec.25.diff_y.HeII.sum.eps}
\caption{Contributions from the \ion{H}{i} (upper panel) and \ion{He}{ii} (lower panel) to the total recombination spectrum for different values of the initial
$y$-parameter. Both the bound-bound and free-bound signals were included. Energy injection was assumed to occur at $z_i=\pot{4}{4}$. The thin red lines
represent the overall negative parts of the signals. We included 25 shells for both \ion{H}{i} and \ion{He}{ii} into our computations. The figures were taken from \citet{Chluba2008a}.}
\label{fig:HI_HeII_y}
\end{figure}

\subsubsection*{Dependence of the cosmological recombination radiation on the $y$-parameter}
\label{RS:sec:prespectrum_y}
Similar to the standard recombination spectrum it is possible to compute the recombination radiation assuming that the ambient CMB radiation field is given by a distorted blackbody, where the distortion is given by a $y$-distortion. 
We first want to address the question how the expected changes in the cosmological recombination radiation depend on the value of the $y$-parameter.
Here the interesting question is if it will be possible to determine the value of the $y$-parameter using the frequency dependence of the CMB spectral distortion.

In Fig.~\ref{fig:HI_HeII_y} we show the results of these computations \citep{Chluba2008a}. In each panel the blue solid line represents the contributions to the normal cosmological recombination spectrum (i.e. $y=0$).
For this case, one can see that the contribution from \ion{He}{ii} is about one order of magnitude smaller than the one from hydrogen.
If we now allow a $y$ distortion with $y=10^{-7}$, then one can see that the contribution from hydrogen has not changed very much. Only a small negative feature, which was completely absent for $y=0$, appeared at $\nu\sim 1200-1300\,$GHz. It is mainly due to high redshift absorption in the Lyman-continuum and the Lyman-series with $n>2$ \citep{Chluba2008a}, and is also visible in the distortion caused by \ion{He}{ii}.
One can also see that already for $y=10^{-7}$ the contribution from \ion{He}{ii} changed more strongly than the one from hydrogen.

This becomes even more apparent, when further increasing the value of the $y$-parameter. Then for both helium and hydrogen the amplitude of the distortion changes several times, where in particular the contribution from \ion{He}{ii} has become comparable to the one from hydrogen.
This is due to the fact that the loops in helium can be run through $\sim 8$ times faster that hydrogen, because of the charge scaling of the atomic transition rates \citep[see][for more detailed explanation]{Chluba2008a}.
Also the negative feature became much more strong, in amplitude even exceeding the Lyman-$\alpha$ distortions from the main recombination epoch.
At low frequencies not only the amplitude of the signal has increased, but also its frequency dependence has changed significantly.
This may allow to determine the value of the $y$-parameter by measuring the frequency-dependent modulation of the CMB spectrum caused due to the presence of atomic species in the early Universe.

\begin{figure}
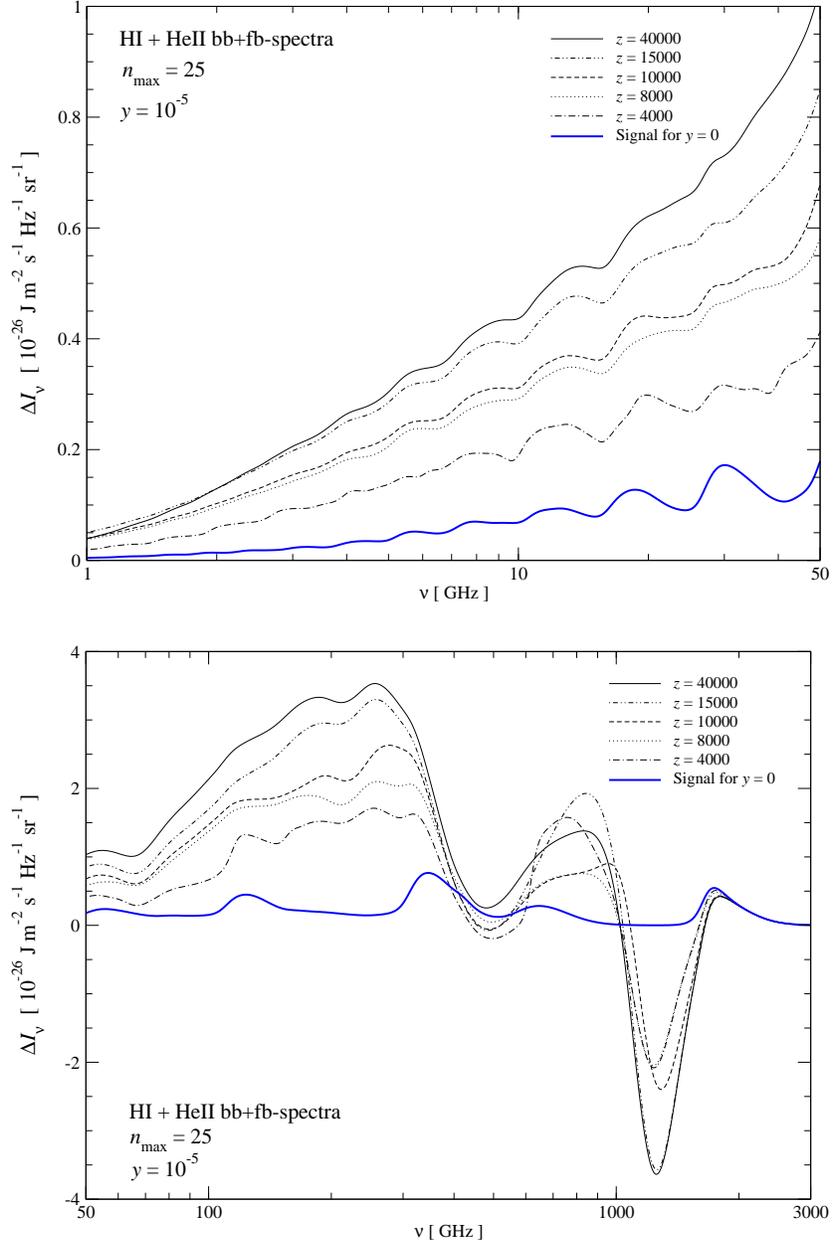

\centering 
\includegraphics[width=0.9\columnwidth]{./HI.HeII.diff_y.low.eps}
\\[5mm]
\includegraphics[width=0.9\columnwidth]{./HI.HeII.diff_y.high.eps}
\caption{Total \ion{H}{i} + \ion{He}{ii} recombination spectra for different
energy injection redshifts. The upper panel shows details of the spectrum at
low, the lower at high frequencies. We included 25 shells  for both \ion{H}{i} and \ion{He}{ii} into our computations. The figures were taken from \citet{Chluba2008a}.}
\label{fig:HI.HeII.diff_z.hl}
\end{figure}
%
Similarly, one can ask how the changes in the cosmological recombination radiation depend on the time of the energy injection. The results of these computations are shown in Fig.~\ref{fig:HI.HeII.diff_z.hl}.
One can see that not only the overall amplitude of the distortion strongly depends on the time of energy injection, but also the shape and number of features changes drastically.
This fact may allow us to understand when the $y$-distortion was actually introduced, and as explained above, at the very least should allow to distinguish {\it pre-} from {\it post-recombinational} $y$-distortions.

\section{Previously neglected physical processes during hydrogen recombination}
\label{RS:sec:processes}
With the improvement of available CMB data also refinements of the
computations regarding the ionization history became necessary, leading to the
development of the widely used {\sc Recfast} code \citep{RS_Seager1999,
  RS_Seager2000, Wong2008}.
However, the prospects with the {\sc Planck} Surveyor have motivated several
groups to re-examine the problem of cosmological hydrogen and helium
recombination,
with the aim to identify previously neglected physical processes that could
affect the ionization history of the Universe at the level of $\gtrsim 0.1\%$.
Such accuracy becomes necessary to achieve the promised precision for the
estimation of cosmological parameters using the observation of the CMB temperature anisotropies and acoustic peaks.
Here we wish to provide an overview of the most important additions in this
context and to highlight some of the previously neglected physical processes \change{during hydrogen recombination}.
Most of them are also important during the epoch of helium recombination
\citep[e.g.][]{RS_HirataI, RS_HirataIII, RS_Jose2007}, but here we focus our discussion on hydrogen only.
The superposition of all effects listed below lead to an ambiguity in
the ionization history during the cosmological hydrogen recombination epoch
that \change{clearly exceeds the level of 0.1\%,  even reaching $\sim 1\%-2\%$ close to the maximum of the Thomson visibility function, where it matters most. All these corrections} therefore should be taken into account in the detailed analysis of future CMB data \citep[for additional overview also see][]{Fendt2009}.
Still the analysis shows that the simple picture, as explained in
Sect.~\ref{RS:sec:Intro} is amazingly stable.

\subsubsection*{Detailed evolution of the populations in the angular momentum sub-states}
\label{RS:sec:pops}
The numerical solution of the hydrogen recombination history and the
associated spectral distortions of the CMB requires the integration of a stiff
system of coupled ordinary differential equations, describing the evolution of
the populations of the different hydrogen levels, with extremely high
accuracy.
Until recently this task was only completed using additional simplifying
assumptions.
Among these the most important simplification is to assume {\it full
  statistical equilibrium}\change{\footnote{i.e. the population of a given level $(n,l)$ is determined by
  $N_{nl} = (2l+1) N_n / n^2$, where $N_n$ is the total population of the
  shell with principle quantum number $n$.}  (SE) within a given shell for
$n>2$.
(for a more detailed comparison of the different approached see
\citet{RS_Jose2006} and references therein).}
However, as was shown in \citet{RS_Jose2006} and \citet{RS_Chluba2007}, this
leads to an overestimation of the hydrogen recombination rate at low redshift
by up to $\sim 3\%-5\%$. This is mainly because during hydrogen recombination
collisions are so much weaker than radiative processes, so that the populations
within a given atomic shell depart from SE. It was also shown that for the
highly excited levels stimulated emission and recombination (see
Fig.~\ref{RS:fig:Rci_Rci_tot}) are important.

\subsubsection*{Induced two-photon decay of the hydrogen 2s-level}
\label{RS:sec:2gamma2s}
\begin{figure}
  \centering 
  \includegraphics[width=0.49\columnwidth]{./RS.Rec_coeffi_lmax.corr.eps}
  \includegraphics[width=0.49\columnwidth]{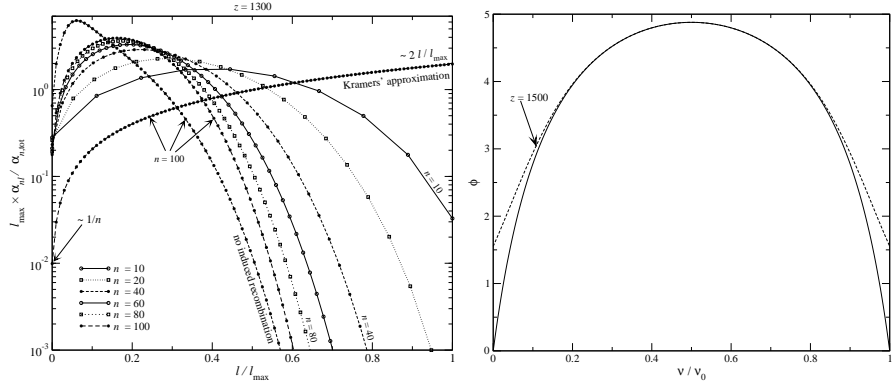}
  \caption{
    {\it Left panel} -- $l$-dependence of the recombination coefficient,
    $\alpha_{nl}$, at $z=1300$ for different shells. The curves have been
    re-scaled by the {\it total} recombination coefficient, $\alpha_{n,\rm
      tot}=\sum_l \alpha_{nl}$, and multiplied by $l_{\rm max}=n-1$ such that
    the 'integral' over $\xi=l/l_{\rm max}$ becomes unity. Also the results
    obtained within the Kramers' approximation, i.e. $\alpha_{nl}^{\rm K}={\rm
      const}\times[2l+1]$, and without the inclusion of stimulated
    recombination for $n=100$ are presented.
    {\it Right panel} -- Two-photon decay profile for the 2s-level in hydrogen: the
    solid line shows the broad two-photon continuum assuming that there is no
    ambient radiation field. In contrast, the dashed line includes the effects
    of induced emission due to the presence of CMB photons at $z=1500$. The
    figures are from \citet{RS_Chluba2006a} and \citet{RS_Chluba2007}.}
\label{RS:fig:Rci_Rci_tot}
\label{RS:fig:2s}
\end{figure}
In the transition of electrons from the 2s-level to the ground state two
photons are emitted in a broad continuum (see Fig. \ref{RS:fig:2s}). Due to
the presence of a large number of CMB photons at low frequencies, stimulated
two-photon emission becomes important when one of the photons is emitted close to the
Lyman-$\alpha$ transition frequency, and, as demonstrated in
\citet{RS_Chluba2006a}, leads to an increase in the effective 2s-1s two-photon
transition rate during hydrogen recombination by more than 1\%.
This speeds up the rate of hydrogen recombination, leading to a maximal change
in the ionization history of $\Delta N_{\rm e}/N_{\rm e}\sim -1.3\%$ at $z\sim
1050$.

\subsubsection*{Re-absorption of Lyman-$\alpha$ photons}
\label{RS:sec:Lya}
The strongest distortion of the CMB blackbody spectrum is associated with the
Lyman-$\alpha$ transition and 2s-1s continuum emission. Due to redshifting
these access photons can affect energetically lower transitions.
%
%
The huge excess of photons in the Wien-tail of the CMB slightly increases the
${\rm 1s}\rightarrow{\rm 2s}$ two-photon absorption rate, resulting in
percent-level corrections to the ionization history during hydrogen
recombination with $\Delta N_{\rm e}/N_{\rm e}\sim +1.9\%$ at $z\sim
1020$ \citep{RS_Kholu2006}.

\subsubsection*{Feedback within the \ion{H}{i}  Lyman-series}
\label{RS:sec:Lynfeedback}
Due to redshifting, all the Lyman-series photons emitted in the transition of
electrons from levels with $n>2$ have to pass through the next lower-lying
Lyman-transition, leading to additional feedback corrections like in the case
of Lyman-$\alpha$ absorption in the 2s-1s two-photon continuum.
However, here the photons connected with Ly$n$ are completely absorbed by the
Ly$(n-1)$ resonance and eventually all Ly$n$ photons are converted into
Lyman-$\alpha$ or 2s-1s two-photon decay quanta.
\change{Also in the computations one has to take into account that the feedback of Ly$n$ photons on the
Ly$(n-1)$ resonance occurs some time after the photon was released. 
For example for Ly$\beta$ to Ly$\alpha$ the feedback happens $\Delta z/z\sim 16\%$ after the emission.}
As shown by \citet{RS_Chluba2007b}, feedback of photons within the \ion{H}{i} Lyman-series leads to a correction in the ionization history of $\Delta N_{\rm e}/N_{\rm e}\sim 0.2\%-0.3\%$ at $z\sim
1050$.
%

\begin{figure}
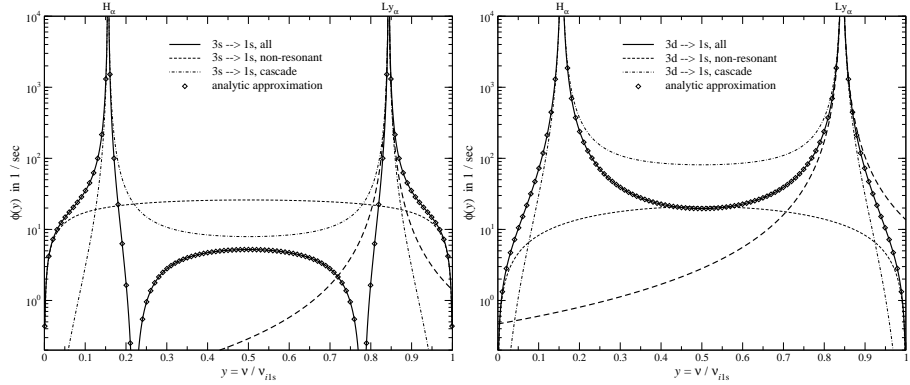

  \centering 
%
  \includegraphics[width=0.49\columnwidth]{./RS.3s.eps}
  \includegraphics[width=0.49\columnwidth]{./RS.3d.eps}
  \caption{Two-photon emission profile for the $3{\rm s}\rightarrow 1{\rm s}$ and $3{\rm
      d}\rightarrow 1{\rm s}$ transition. The non-resonant, cascade and
    combined spectra are shown as labeled. Also we give the analytic
    approximation as given in \citet{RS_Chluba2007c} and show the usual
    Lorentzian corresponding to the Lyman-$\alpha$ line (long dashed). 
%
%
    The figure is from \citet{RS_Chluba2007c}.}
\label{RS:fig:3s3d}
\end{figure}
\subsubsection*{Two-photon transitions from higher levels}
\label{RS:sec:2ngamma2s}
One of the most promising additional corrections to the ionization history is
due to the two-photon transition of highly excited hydrogen states to the
ground state as proposed by \citet{RS_Dubrovich2005}.
The estimated correction was anticipated to be as large as $\sim 5\%$ very
close to the maximum of the Thomson visibility function, and therefore should
have had a large impact on the theoretical predictions for the CMB power
spectra.
It is true that in the extremely low density plasmas the cascade of permitted
transitions (for example the chain 3s$\rightarrow$2p$\rightarrow$1s) goes
unperturbed and might be considered as two photon process with two resonances
\citep{RS_Goeppert}. In addition there is a continuum, \change{analogues to 2s-1s decay
spectrum, and an interference term between resonant contributions} and this weak continuum (see
Fig.  \ref{RS:fig:3s3d} and \citet{RS_Chluba2007c}).
However, the estimates of \citet{RS_Dubrovich2005} only included the
contribution to the two-photon decay rate coming from the two-photon continuum, \change{which is
due to virtual transitions}, and as \change{for example} shown in \citet{RS_Chluba2007c} in
particular the interference between \change{resonant and non-resonant contributions}
plays an important role in addition.
\change{This results in deviations of the line emission profiles from the normal Lorentzian shape, which are caused by quantum mechanical aspects of the problem and are most strong in the distant damping wings (e.g. see Fig.~\ref{RS:fig:3s3d}).}

\begin{figure}
\centering 
\includegraphics[width=0.7\columnwidth]{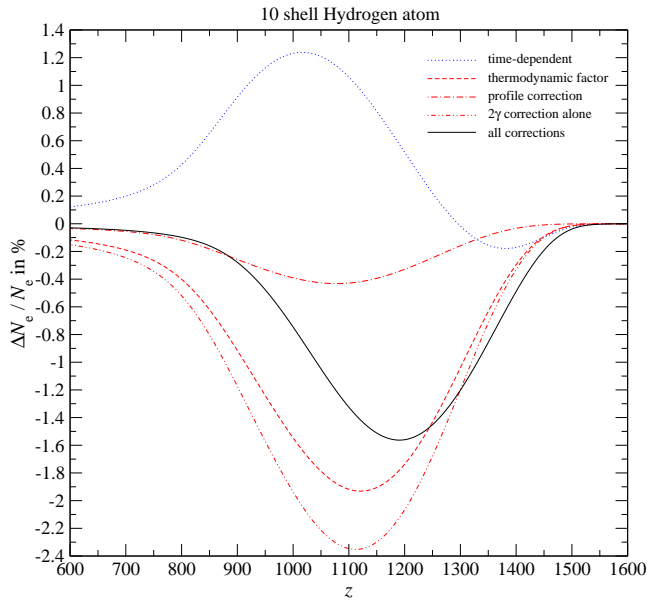}
\caption
{Changes in the free electron fraction: separate contributions due to the {\it time-dependent} correction, the {\it thermodynamic factor} and the {\it shape} of the profile. The figure was taken from \citet{Chluba2009}.}
\label{fig:DN_N.final}
\end{figure}
\change{
Furthermore, as for example pointed out by \citet{RS_Chluba2007c} the full problem has to include aspects of the radiative transfer in the main resonances, since some significant fraction of photons are also escaping from within a few ten to hundred Doppler width of the line centers. In addition, at the percent-level even in the very distant damping wings (i.e. $10^2-10^3$ Doppler width away from the line center) radiative transfer is still important, leading to additional re-absorption of photons before they finally escape.
Recently this problem was considered in detail by \citet{Chluba2009} for the 3d-1s and 3s-1s two-photon transitions.
In their analysis, three independent sources of corrections were identified (we will discuss the other two processes below), showing that the total modification coming from purely quantum mechanical aspects of the problem  lead to a change in the free electron number of $\Delta N_{\rm e}/N_{\rm e}\sim -0.4\%$ at $z\sim 1100$ (see Fig.~\ref{fig:DN_N.final} for more detail).
}

\subsubsection*{Time-dependent aspects in the emission and absorption of Lyman $\alpha$ photon}
\label{RS:sec:time}
One of the key ingredients for the derivation of the escape probability in the Lyman $\alpha$ resonance using the Sobolev approximation \citep{Sobolev1960} is the {\it quasi-stationarity} of the line transfer problem. However, as shown recently \citep{Chluba2009a, Chluba2009} at the percent-level this assumption is not justified during the recombination of hydrogen, since (i) the ionization degree, expansion rate of the Universe and Lyman $\alpha$ death probability change over a characteristic time $\Delta z/z\sim 10\%$, and (ii) because a significant contribution to the total escape probability is coming from photons emitted in the distant wings (comparable to $10^2-10^3$ Doppler width) of the Lyman $\alpha$ resonance.
Therefore one has to include {\it time-dependent aspects} in the emission and absorption process into the line transfer problem, leading to a delay of recombination by $\Delta N_{\rm e}/N_{\rm e}\sim +1.2\%$ at $z\sim 1000$ (see Fig.~\ref{fig:DN_N.final} for more detail).

\subsubsection*{Thermodynamic asymmetry in the Lyman $\alpha$ emission and absorption profile}
\label{RS:sec:thermo}
Knowing the shape of the Lyman $\alpha$ line emission profile\footnote{It is usually assumed to be given by a Voigt profile.} and applying the {\it detailed balance principle}, one can directly obtain an expression for the line absorption profile. 
With this one finds that there is a {\it frequency-dependent asymmetry} between the line emission and absorption profile, which becomes strongest at large distances (beyond $10^2-10^3$ Doppler width) from the line center.
This asymmetry is given by a {\it thermodynamic correction factor} \citep{Chluba2009}, which has an exponential dependence on the {\it detuning} from the line center, i.e. $f_\nu\propto \exp(h[\nu-\nu_\alpha]/kT_\gamma)$, where $\nu_\alpha$ is the transition frequency for the Lyman alpha resonance.
Usually this factor can be neglected, since for most astrophysical problems the main contribution to the number of photons is coming from within a few Doppler width of the line center, where the thermodynamic factor indeed is very close to unity.
However, as mentioned above, in the Lyman $\alpha$ escape problem during hydrogen recombination also contributions from the very distant damping wings are important, so that there $f_\nu\neq1$ has to be included.

In the normal $'1+1'$ photon picture for the line emission and absorption process $f_\nu$ has no direct physical interpretation. It is simply the result of thermodynamic requirements necessary to perserve a blackbody spectrum at all frequencies from the line center in the case of full thermodynamic equilibrium. 
However, as explained by \citet{Chluba2009}, in the two photon picture $f_\nu$ is due to the fact that in the recombination problem the photon distribution for transitions from the 2p-state towards higher levels or the continuum is given by the CMB blackbody radiation.
For example, once an electron reached the 2p-state by the absorption of the Lyman $\alpha$ photon $\gamma_1$, it will only be able to be further excited, say to the 3d-level, by the aid of a Balmer $\alpha$ photon $\gamma_2$ from the ambient CMB radiation field. If the energy of the photon $\gamma_1$ was initially a bit smaller than the Lyman $\alpha$ frequency, then this lack of energy has to be compensated by the photon $\gamma_2$, since due to energy conservation $\nu_{\gamma_1}+\nu_{\gamma_2}$ should equal the transition frequency to the third shell, $\nu_{31}$.
Because during hydrogen recombination blue-ward of the Balmer $\alpha$ resonance there are exponentially fewer photons in the CMB than at the line center, the efficiency of Lyman $\alpha$ absorption is exponentially smaller in the red wing of the Lyman $\alpha$ resonance.
With a similar argument, the absorption efficiency is exponentially larger in the blue wing of the Lyman $\alpha$ resonance.
This process leads to a $\sim 10\%$ increase in the Lyman $\alpha$ escape probability, and hence speeds hydrogen recombination up. \citet{Chluba2009} obtained $\Delta N_{\rm e}/N_{\rm e}\sim -1.9\%$ at $z\sim 1100$ (see Fig.~\ref{fig:DN_N.final} for more detail).

One should also mention that this large change in the escape probability of Lyman $\alpha$ photons will directly translate into similar changes in the amplitude of the Lyman $\alpha$ line, although the electron fraction was affected by much less.
Also, the shape of the low-frequency distortion from highly excited level will be affected by this process, so that the recombination spectrum in principle should allows us to understand the details in the dynamics of hydrogen recombination.

\begin{figure}
\centering
\includegraphics[width=0.7\columnwidth]{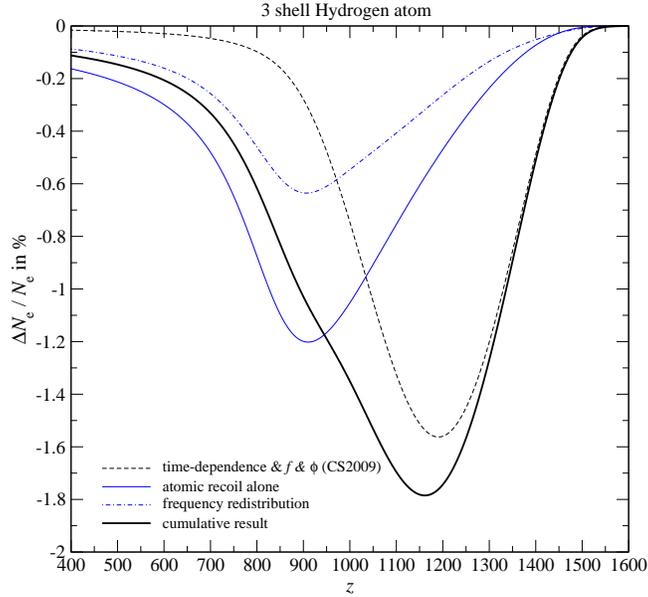}
\caption
{Changes in the free electron fraction due to partial frequency redistribution, including {\it atomic recoil}, and {\it Doppler-broadening} and {\it boosting}. The curve labeled 'CS2009' is the total result from Fig.~\ref{fig:DN_N.final}. The figure was taken from \citet{Chluba2009b}.}
\label{fig:DN_N.final.diff}
\end{figure}
\subsubsection*{Partial frequency redistribution and its effect on the recombination history}
\label{RS:sec:partial}
The other key ingredients for the derivation of the escape probability in the Lyman $\alpha$ resonance using the Sobolev approximation \citep{Sobolev1960} is the assumption that every line scattering leads to a {\it complete-redistribution} of photons over the whole line profile. 
It is clear that this assumption is not very accurate, since in each scattering photons will only be redistributed by $\Delta\nu/\nu \sim 10^{-5}-10^{-4}$, where the redistribution is related to the Doppler motion of the atom \citep{Hummer1962, RS_Rybicki1993}. This again is due to the absence of collisions, since without them a complete redistribution of photons over the Lyman $\alpha$ line profile can only occur when the 2p electron is further excited towards higher levels, forgetting its history on the way. The latter process is related to an absorption event rather than a line scattering. However, during hydrogen recombination the probability for this is about $10^3 -10^4$ smaller than the scattering rate \citep[e.g. see][]{RS_Chluba2007c}, so that a complete redistribution becomes rather unlikely in particular when going to the distant line wings, where the total scattering rate is significantly smaller than in the line center \citep[also see explanations in][]{RS_Chluba2007c, RS_HirataI, Chluba2009a}.

This has lead several groups to consider the frequency redistribution of Lyman $\alpha$ photons in this problem more carefully. Since the Lyman $\alpha$ scattering rate is huge during hydrogen recombination one can use a Fokker-Planck approximation for the redistribution function \citep[e.g. see][]{Rybicki2006}.
Here three processes are important: (i) atomic recoil\footnote{This term was first introduced by \citet{Basko1978, Basko1981}}, (ii) Doppler
boosting, and (iii) Doppler broadening. All three physical processes
are also well-known in connection with the Kompaneets
equation which describes the repeated scattering of photons by
free electrons. 

Atomic recoil leads to a {\it systematic drift} of photons
towards lower frequencies after each resonance scattering.
This allows some additional photons to escape from the Lyman
$\alpha$ resonance and thereby speeds hydrogen recombination up, as
demonstrated by \citet{Grachev2008} and others \citep{Chluba2009b, Hirata2009}.
However, the processes (ii) and (iii) were neglected in the analysis of \citet{Grachev2008}.
As recently shown by \citet{Chluba2009b}, Doppler boosting acts in the opposite direction as atomic recoil and therefore should slow recombination down, while the effect of Doppler broadening can lead to both an increase in the photons escape probability of a decrease, depending on the initial frequency of the photons \citep[see][for more detailed explanation]{Chluba2009b}. 
The overall correction to the recombination history due to processes (i)-(iii) is dominated by the one caused by  atomic recoil effect, and amounts to $\Delta N_{\rm e}/N_{\rm e}\sim -0.6\%$ at $z\sim 900$ (see Fig.~\ref{fig:DN_N.final.diff} for more detail).
The results of the computations by \citet{Chluba2009b} seem to be in very good agreement with those from \citet{Hirata2009}.

\begin{figure}
\centering 
\includegraphics[width=0.75\columnwidth]{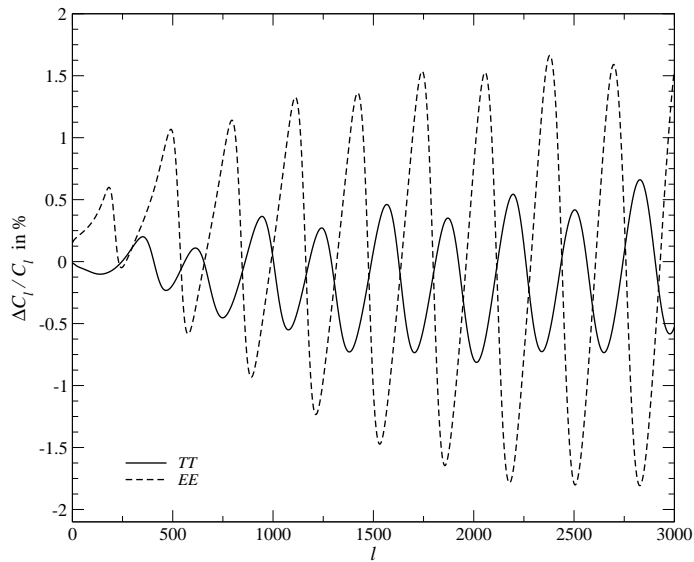}
\caption
{Changes in the CMB temperature and polarization power spectra caused by the cumulative effect of {\it partial frequency redistribution}, the {\it time-dependent correction}, the {\it thermodynamic factor}, and the correction due to the {\it shape} of the effective emission profile. The figure was taken from \citet{Chluba2009b}.}
\label{fig:DCl}
\end{figure}
\change{
Finally, in Fig.~\ref{fig:DCl} we show as an example the cumulative effect on the CMB temperature and polarization power spectra caused by the corrections in the ionization history due to {\it partial frequency redistribution}, the {\it time-dependent correction}, the {\it thermodynamic factor}, and the correction due to the {\it shape} of the effective emission profile.
In particular the associated changes in the EE power spectrum are impressive, reaching
peak to peak amplitude $\sim 2\% - 3\%$ at $l \geq 1500$. It
will be important to take these corrections into account for the
analysis of future CMB data.}

\subsubsection*{Additional processes during hydrogen recombination}
\label{RS:sec:add}
There are a few more processes that here we only want mention very briefly (although with this the list is not meant to be absolutely final or complete). \citet{Hirata2008} also included the {\it two-photon decays from higher levels} in hydrogen and 2s-1s {\it Raman scattering}. The former lead to an additional speed up of hydrogen recombination at the level of $\Delta N_{\rm e}/N_{\rm e}\sim 0.1\%-0.3\%$, while the Raman process leads to an additional delay of recombination by $\Delta N_{\rm e}/N_{\rm e}\sim 0.9\%$ at $z\sim 900$.

The effect of {\it electron scattering} during hydrogen recombination was also recently investigated by \citet{Chluba2009b} using a Fokker-Planck approach. This approximation for the frequency redistribution function may not be sufficient towards the end of hydrogen recombination, but in the overall correction to the ionization history was very small close the maximum of the Thomson visibility function, so that no big difference are expected when more accurately using a scattering Kernel-approach.

One should also include the small re-absorption of photons from the 2s-1s two-photon continuum close to the Lyman $\alpha$ resonance, where our estimates show that this leads to another $\Delta N_{\rm e}/N_{\rm e}\sim 0.1\%$ correction.
Also the feedback of helium photons on hydrogen recombination poses an interesting problem, but the changes in the ionization history are negligible \citep{Chluba2009c}.

\subsection{Towards a new recombination code}
\label{RS:sec:reccode}
The list of additional processes that have been studied in connection with the cosmological recombination problem is already very long. 
However, it seems that most of the important terms have been identified, so that now it is time to think about the inclusion of all these processes in a {\it new cosmological recombination code}, which then can be used in the analysis of CMB data as will become available with {\sc Planck} soon.
The important steps towards this new code will be (i) to {\it cross validate} all the discussed corrections by independent groups/methods, and (ii) to develop a scheme that is sufficiently {\it fast} and {\it precise} and still captures all the important corrections.

The first step is particularly important, since at percent-level accuracy it is very easy to make mistakes, even if they are only due to numerics.
For the second point the problem is that one run of the full recombination code will likely take far too long\footnote{In the current implementation our code would take of the order of a week for one cosmology.} to be useful for parameter estimation from CMB data.
To solve this problem three strategies could be possible: (a) one can find appropriate {\it fudge functions} to mimic the recombination dynamics using {\sc Recfast}; (b) one can try to find an approximate, physically motivated representation of the problem; or (c) one can simply tabulate the outputs of the full recombination code for different cosmologies and then interpolate on the obtained grid of models.

In connection with this we would like to advertise the work of \citet{Fendt2009},  leading to the development of {\sc Rico}\footnote{http://cosmos.astro.uiuc.edu/rico}, which uses
multi-dimensional polynomial regression to accurately represent the dependence of the free electron fraction on redshift and the cosmological parameters.
{\sc Rico} is both very fast and accurate, and can be {\it trained} using any available recombination code. This feature in addition makes it very interesting in connection with code comparisons and when looking for more approximate, physically motivated representations of the problem.
Once we finished our final recombination code we plan on providing the training sets for {\sc Rico}, so that it then can be used in the data analysis in connection with {\sc Planck}.

\section{Conclusions}
\label{RS:sec:conclusion}
It took several decades until measurements of the CMB temperature fluctuations
became a reality.
After {\sc Cobe} the progress in experimental technology has accelerated by
orders of magnitude. 
Today CMB scientists are even able to measure $E$-mode polarization, and the
future will likely allow us to access the $B$-mode component of the CMB in
addition.
Similarly, one may hope that the development of new technologies will render
the consequences of the discussed physical processes observable.
Therefore, also the photons emerging during the epochs of cosmological
recombination could open another way to refine our understanding of the
Universe.
As we illustrated in this contribution, by observing the CMB spectral
distortions from the epochs of cosmological recombination we can in principle
directly measure cosmological parameters like the value of the CMB monopole
temperature, the specific entropy, and the pre-stellar helium abundance, {\it
not suffering} from limitations set by {\it cosmic variance}.
Furthermore, we could directly test our detailed understanding of the
recombination process using {\it observational data}.
{It is also remarkable that the discussed CMB signals are coming from
redshifts $z\sim 1300-1400$ for hydrogen, $z\sim 1800-1900$ for neutral
helium, and $z\sim 6000$ for \ion{He}{ii}. This implies that by observing
these photons from recombination we can actually look beyond the last
scattering surface, i.e. before bulk of the CMB temperature anisotropies were actually formed.}
To achieve this task, {\it no absolute measurement} is necessary, but one only
has to look for a modulated signal at the $\sim\mu$K level, with typical
amplitude of $\sim 10-30\,$nK and $\Delta \nu/\nu\sim 0.1$, where this signal
{in principle} can be predicted with high accuracy, yielding a {\it
spectral template} for the full cosmological recombination spectrum, also
including the contributions from helium.
\change{The combination of both {\it spectral} and {\it spatial} fluctuation in the CMB blackbody temperature may therefore eventually allow us to perform purely CMB based parameter estimations, yielding competitive constraints on the Universe we live in.}

\change{And finally, if something {\it unexpected} happened {\it during} or {\it before} the recombination epoch, then this may leave observable traces in the cosmological recombination radiation.
We have illustrated this statement for the case of energy injection in the pre-recombinational epoch (Sect.~\ref{RS:sec:prespectrum}), but also if something unexpected occurred during the recombination of hydrogen, e.g leading to {\it delayed recombination} \citep{Peebles2000}, then this should leave signatures in the cosmological recombination radiation, affecting {\it not only} the shape of the Lyman $\alpha$ distortion, but also the low frequency part of the recombination spectrum. 
This might help us to place tighter constraints on the thermal history of our Universe and the physics of cosmological recombination.
}



\end{document}